\documentclass[twocolumn]{aastex631}

\usepackage{tabularx}
\usepackage{amssymb}

\usepackage{amsmath}

\begin{document}

\title{From pre-transit to post-eclipse: investigating the impact of 3D temperature, chemistry, and dynamics on high-resolution emission spectra of the ultra-hot Jupiter WASP-76b}

\author[0000-0003-3191-2486]{Joost P. Wardenier}
\altaffiliation{joost.wardenier@umontreal.ca}
\affiliation{Institut Trottier de Recherche sur les Exoplanètes, Université de Montréal, Montréal, Québec, H3T 1J4, Canada}

\author[0000-0001-9521-6258]{Vivien Parmentier}
\affiliation{Université Côte d’Azur, Observatoire de la Côte d’Azur, CNRS, Laboratoire Lagrange, Nice, France}

\author[0000-0002-3052-7116]{Elspeth K. H. Lee}
\affiliation{Center for Space and Habitability, University of Bern, Gesellschaftsstrasse 6, CH-3012 Bern, Switzerland}

\author[0000-0002-2338-476X]{Michael R. Line}
\affiliation{School of Earth $\&$ Space Exploration, Arizona State University, Tempe AZ 85287, USA}



\begin{abstract}
High-resolution spectroscopy has provided a wealth of information about the climate and composition of ultra-hot Jupiters. However, the 3D structure of their atmospheres makes observations more challenging to interpret, necessitating 3D forward-modeling studies. In this work, we model phase-dependent thermal emission spectra of the archetype ultra-hot Jupiter WASP-76b to understand how the line strengths and Doppler shifts of Fe, CO, H$_2$O, and OH evolve throughout the orbit. We post-process outputs of the SPARC/MITgcm global circulation model with the 3D Monte-Carlo radiative transfer code gCMCRT to simulate emission spectra at 36 orbital phases. We then cross-correlate the spectra with different templates to obtain CCF and $K_{\text{p}}$--$V_{\text{sys}}$ maps. For each species, our models produce consistently negative $K_{\text{p}}$ offsets in pre- and post-eclipse, which are driven by planet rotation. The size of these offsets is similar to the equatorial rotation velocity of the planet. Furthermore, we demonstrate how the weak vertical temperature gradient on the nightside of ultra-hot Jupiters mutes the absorption features of CO and H$_2$O, which significantly hampers their detectability in pre- and post-transit. We also show that the $K_{\text{p}}$ and $V_{\text{sys}}$ offsets in pre- and post-transit are not always a measure for the line-of-sight velocities in the atmosphere. This is because the cross-correlation signal is a blend of dayside emission and nightside absorption features. Finally, we highlight that the observational uncertainty in the known orbital velocity of ultra-hot Jupiters can be multiple km/s, which makes it hard for certain targets to meaningfully report absolute $K_{\text{p}}$ offsets. 
\end{abstract}



\label{sec:intro}
\section{Introduction}

Owing to their short orbital periods, high equilibrium temperatures, and puffy atmospheres, ultra-hot Jupiters (UHJs; \citealt{Parmentier2018,Bell2018,Arcangeli2018}) are prime targets for atmospheric characterization via ground-based high-resolution spectroscopy. With the arrival of a host of new instruments including ESPRESSO \citep{Pepe2021}, \mbox{IGRINS} \citep{Mace2018}, \mbox{IGRINS-2} \citep{Oh2024}, \mbox{MAROON-X} \citep{Seifahrt2018}, NIRPS \citep{Artigau2024}, CRIRES+ \citep{Dorn2023}, KPIC \citep{Delorme2021}, and GHOST \citep{Kalari2024}, we are now able to study the climate, chemical composition, and formation history of UHJs in unprecedented detail. Yet, one aspect that complicates the interpretation of the spectra of these planets is the inherent 3D structure of their atmospheres. 

UHJs are tidally locked, such that they have a permanent dayside ($T$ $\sim$ 3000 K) and a permanent nightside ($T$ $\sim$ 1000 K). The dayside temperature is so high that molecules like H$_2$ and H$_2$O are subject to thermal dissociation, resulting in significant amounts of H, H$^-$, and OH (\citealt{Parmentier2018,Yan2021,Nugroho2021}). Furthermore, a large fraction of metals become ionized (\citealt{Hoeijmakers2019,Merritt2021,Prinoth2024}). The absorption of intense stellar radiation by metal species, and potentially by TiO and VO, also causes the dayside temperature profile to be strongly inverted, such that temperature increases with altitude (\citealt{Fortney2008,Haynes2015,Pino2020}). On the dark nightside, however, the temperature profile should be non-inverted. Also, the lower nightside temperature allows for H$_2$ and H$_2$O to persist, while causing refractory species to condense into clouds (\citealt{Helling2021,Komacek2022}).

When it comes to the signature of ``3D-ness'' in the transmission spectra of UHJs, a relatively coherent picture has emerged. Absorption lines of species that probe the dayside (e.g., metals and CO) tend to show an increasing blueshift during the transit (\citealt{Ehrenreich2020,Kesseli2022,Savel2023,Wardenier2023,Pelletier2023,Prinoth2023,Beltz2023,Simonnin2024,Seidel2025}). This is a consequence of the strong day-night contrast of the planet in combination with tidally locked rotation. The day-to-night flow produces an additional negative velocity offset of a few km/s. As a result, the detection peaks of these ``dayside'' species occur at smaller $K_{\text{p}}$ and $V_{\text{sys}}$ values compared to the expected planet position in the $K_{\text{p}}$--$V_{\text{sys}}$ map of the observation (\citealt{Wardenier2023}). For species that mainly probe the nightside of the planet (e.g., H$_2$O), the Doppler shifts of the absorption lines can either become more negative or more positive during the transit (\citealt{Wardenier2023,Wardenier2024}), depending on the atmospheric drag strength. On top of this behavior, asymmetries in scale height (i.e., temperature), cloud cover, or the degree of ionization between the planet's morning and evening terminator can cause the absorption line strengths to vary substantially during the transit (\citealt{Wardenier2021,Savel2022,Gandhi2022,Prinoth2023}). Typically, UHJs have a hotter morning limb due to an eastward circulation pattern.

With regard to emission spectroscopy, a number of studies have simulated high-resolution thermal emission spectra of hot and ultra-hot Jupiters based on 3D global circulation models (GCMs) (\citealt{Zhang2017,Harada2019,Malsky2021,Beltz2020,Beltz2022,Lee2022b,vanSluijs2022,Beltz2024}). The common denominator between many of these works is the finding that the planet's 3D temperature structure has a more profound effect on spectral line shapes than in transmission. In a transmission spectrum, the line strength is determined by the \emph{altitude difference} between the pressure probed by the line core and the pressure probed by the continuum (e.g., \citealt{Wardenier2022,Wardenier2023}). In an emission spectrum, however, the line strength is directly set by the \emph{thermal flux difference}\footnote{In thermal emission, the flux difference between a line core at wavelength $\lambda$ and the adjacent continuum would be proportional to $B_\lambda(T_{\text{core}})$ -- $B_\lambda(T_{\text{cont}})$, with $B_\lambda$ the Planck function and $T_{\text{core}}$ and $T_{\text{cont}}$ the temperatures probed by the line core and the continuum, respectively.} between these two pressures (e.g., \citealt{vanSluijs2022}). If the line core probes a higher temperature than the continuum (due to a thermal inversion), the spectrum contains \emph{emission} lines. If the temperature profile is non-inverted, the corresponding emission spectrum will contain \emph{absorption} lines. Because an emission spectrum is the integrated flux over the entire planet disk (which contains different portions of the dayside and nightside at different orbital phases), it will originate from regions with different vertical temperature gradients, giving rise to line shapes that cannot be fully captured by 1D atmospheric models (e.g., \citealt{Beltz2020}).

The GCM studies also predicted how the Doppler shifts of emission spectra vary in the planet rest frame as a function of orbital phase. In principle, the Doppler shifts of a tidally locked planet should follow a ``quasi-sinusoidal'' behavior (\citealt{Zhang2017,Harada2019,Malsky2021,Beltz2022}) as the dayside rotates towards the observer in pre-eclipse (causing a blueshift) and away from the observer in post-eclipse (causing a redshift). Just like in transmission, these anomalous Doppler shifts should naturally lead to peak offsets in $K_{\text{p}}$--$V_{\text{sys}}$ maps obtained from UHJ emission observations. Furthermore, the combination of tidally locked rotation and the 3D temperature structure should cause line strengths to vary substantially along the orbit. 

Indeed, numerous emission studies of UHJs have reported species whose cross-correlation signals deviate from the known planetary $K_{\text{p}}$ and/or $V_{\text{sys}}$ values, or for which the signal strengths change between different orbital phases (e.g., pre-eclipse vs post-eclipse). This includes observations of Fe, CO, and H$_2$O on \mbox{WASP-76b} \citep{Yan2023,costasilva2024}, CO, H$_2$O, OH, and various metal species on \mbox{WASP-121b} \citep{Smith2024,Hoeijmakers2022,Pelletier2024,Bazinet2025}, CO, H$_2$O, and OH on \mbox{WASP-18b} \citep{Brogi2023,Yan2023}, Fe and CO on \mbox{WASP-189b} \citep{Yan2020,Yan2022_CO,Lesjak2024}, Fe, CO, and TiO on \mbox{WASP-33b} \citep{Nugroho2017,Cont2021,Cont2022,Herman2022,Yan2022_CO,vanSluijs2022,Finnerty2023,Mraz2024}, Fe, Fe+ and Cr on KELT-20b \citep{Yan2022,Borsa2022}, and Fe on \mbox{KELT-9b} \citep{Pino2020,Pino2022}. 

\begin{deluxetable}{ccccc}[t!]

{\color{black}{

    \tablewidth{0pt} 
    \tablecaption{Overview of some of the parameters of the WASP-76b GCMs described in Section \ref{subsec:gcm_methods} (see Figs. \ref{fig:globes_abunds} to \ref{fig:globes_vlos} for plots of the temperature structures, abundances and line-of-sight velocities).}
    \tablehead{
        \multicolumn{1}{c}{Parameter} &
        \multicolumn{1}{c}{Value} &
    }
     \label{tab:gcm_parameters}
      \startdata
        Orbital period & 1.5637$\times$$10^5$ s (1.81 days) \\
        Pressure range & 200 -- 2$\times$10$^{-6}$ bar \\
        Radius at bottom & 1.3038$\times$$10^8$ m (1.824 R$_{\rm Jup}$) \\
        Gravity & 7.6 m/s$^2$ \\  
        Horizontal resolution & C32 \\
        Vertical resolution & 53 layers \\
        Metallicity and C/O & 1 $\times$ solar \\
        Drag timescale & $\{\infty, 10^5 \ \text{s}, 10^4 \ \text{s}\}$ \\
        Radiative transfer & non-grey (see \citealt{Kataria2013}) \\
    \enddata

}}
\vspace{-15pt}
\end{deluxetable}

The aim of this work is to build on previous GCM modeling studies and further investigate how the ``3D-ness'' of UHJs impacts high-resolution observables in thermal emission, such as peak offsets in $K_{\text{p}}$--$V_{\text{sys}}$ maps and phase-dependent signal strengths. We will also assess the detectability of nightside spectra of UHJs. The structure of this manuscript is as follows. Section \ref{sec:methods} summarizes our GCMs and 3D radiative-transfer framework. In Section \ref{sec:results}, we present and discuss our results. Finally, Section \ref{sec:conclusion} provides a summary and conclusion.   


\section{Methods}
\label{sec:methods}

\subsection{Global circulation models}
\label{subsec:gcm_methods}

In this work, we focus on three 3D atmospheric models of the canonical UHJ WASP-76b ($T_{\text{eq}}$ $\sim$ 2200 K; \citealt{West2016}), generated with the SPARC/MITgcm (\citealt{Showman2009}). These are the same models that were used in \citet{Wardenier2021,Wardenier2023} to study the phase-dependent behavior of the planet's \emph{transmission} spectrum at high resolution. The SPARC/MITgcm is a state-of-the-art, non-grey GCM that has been used to explore many aspects of the atmospheric physics and chemistry of hot gas giants over the last fifteen years (e.g., \citealt{Fortney2010,Showman2013,Kataria2013,Parmentier2016,Parmentier2018,Steinrueck2020,Tan2024,Roth2024}).

Table \ref{tab:gcm_parameters} provides an overview of the most important parameters of our WASP-76b models. For further information about how the GCMs were run, we refer the reader to \citet{Parmentier2018} and \citet{Wardenier2021}. In contrast to some previous works that focused on cooler hot Jupiters (\citealt{Harada2019,Malsky2021}), we do not consider the effects of clouds. In the context of UHJs, such an approach is justified as their daysides, which emit the bulk of the planet's thermal flux, are expected to be completely cloud free (e.g., \citealt{Komacek2022}). The setup of each of the WASP-76b models is identical. The only parameter that is varied is the (uniform) drag timescale $\tau_{\text{drag}}$, such that we obtain atmospheres with no drag \mbox{($\tau_{\text{drag}} \rightarrow \infty$)}, weak drag \mbox{($\tau_{\text{drag}} = 10^5$ s)}, and strong \mbox{drag ($\tau_{\text{drag}} = 10^4$ s)}, respectively. The drag timescale represents the typical time it takes for a parcel of air to lose a significant fraction of its kinetic energy. It accounts for a number of physical processes, such as turbulent mixing (\citealt{Li2010}), Lorentz-force braking of ionized winds in the planet's magnetic field (\citealt{Perna2010}), or Ohmic dissipation (\citealt{Perna2010a}). From an observational perspective, $\tau_{\text{drag}}$ is an important parameter as it governs the effective wind speeds, and thereby the efficiency of heat redistribution, as well as the Doppler shifts observed in the associated spectra. While the drag-free model has a super-rotating equatorial jet \citep{Wardenier2021}, jet formation in the other two models is suppressed, leading to a wind profile that is mainly composed of a day-to-night flow. 

Before calculating the thermal emission spectra associated with our WASP-76b models, we first map the GCM outputs onto a 3D grid with altitude (instead of pressure) as a vertical coordinate. The details of this procedure are described in Section 2.1 in \citet{Wardenier2023}. Because the dayside has a larger scale height than the nightside due to its higher temperature and lower mean-molecular weight (resulting from the thermal dissociation of H$_{\text{2}}$), we extrapolate the nightside down to arbitrarily low pressures to make sure that the atmosphere reaches the same altitude on both hemispheres. The contribution from these pressures to the emitted flux is virtually zero. After binning and extrapolating, our models consist of 32$\times$64$\times$58 $\sim$ 10$^5$ cells in latitude, longitude, and altitude, respectively. 

\subsection{Computing emission spectra with gCMCRT}

\begin{figure}
\vspace{-10pt}
{\hspace{-20pt} \includegraphics[width=0.54\textwidth]{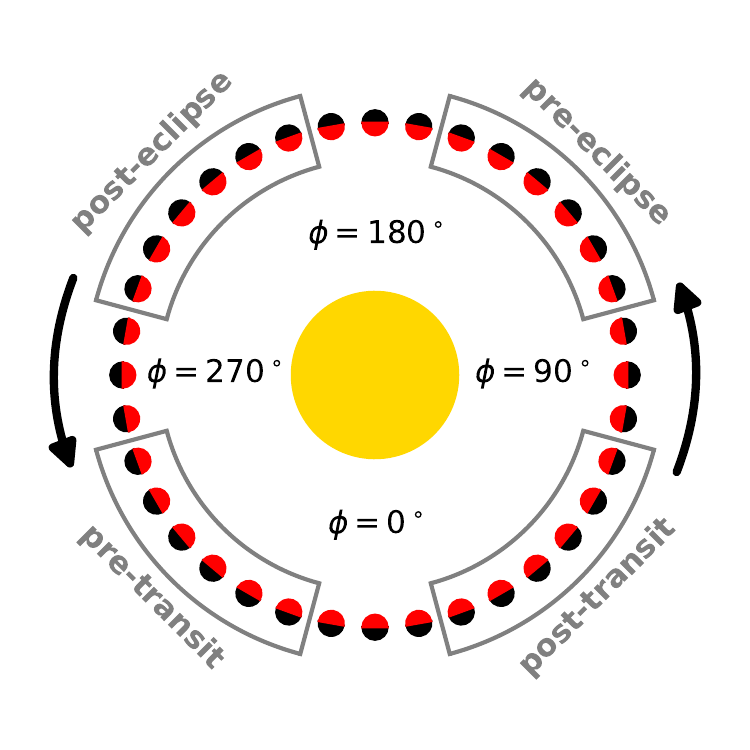}}
\vspace{-35pt}
\caption{Schematic overview of the orbit of a tidally locked planet, with the dayside in red and the nightside in black (sizes and distances not to scale). The transit corresponds to phase $\phi = 0^\circ$, while the secondary eclipse occurs at $\phi = 180^\circ$. In this work, we simulate the emission spectra of our models at 36 orbital phases at $10^\circ$ intervals. The grey boxes indicate the parts of the orbit which we define as the pre-transit \mbox{($285^\circ$ -- $345^\circ$)}, post-transit ($15^\circ$ -- $75^\circ$), pre-eclipse ($105^\circ$ -- $165^\circ$), and post-eclipse ($195^\circ$ -- $255^\circ$), respectively.}
\label{fig:schematic_orbit}
\end{figure}

We use the 3D Monte-Carlo radiative transfer code gCMCRT \citep{Lee2022} to simulate phase-dependent thermal emission spectra for each of the GCM outputs. As shown in Fig. \ref{fig:schematic_orbit}, we calculate spectra at 36 orbital phases $\phi$, which are separated by 10-degree intervals. Because the planet is tidally locked, different portions of the dayside and nightside are in view at different orbital phases. We do not consider the effects of clouds or multiple scattering in the radiative transfer to avoid overcomplicating the model. 

The full inner workings of gCMCRT are described in \citet{Lee2022}, so we will just summarize the most important steps below. For a certain wavelength $\lambda$, each atmospheric cell $i$ has a luminosity $L_{i}$ given by 

\begin{equation} \label{eq:cell_luminosity2}
L_{i}(\lambda) = 4 \pi \rho_i V_i \tilde{\kappa}_{i}(\lambda, v_{\text{los}}) B_{\lambda}(T_i), 
\end{equation}

\noindent where $\tilde{\kappa}_{i}$ is a Doppler-shifted opacity to account for the fact that the emitting cell is moving at a velocity $v_{\text{los}}$ with respect to the observer (due to planet rotation and the wind profile; see \citealt{Wardenier2021}):

\begin{equation} \label{eq:doppler_shift}
\tilde{\kappa}_i(\lambda, v_{\text{los}}) = \kappa_i(\lambda_{\text{eff}}), \ \ \text{with} \ \lambda_{\text{eff}} = \lambda \bigg[ 1-\frac{v_{\text{los}}}{c} \bigg].
\end{equation}

\noindent In the above equations, $\rho_i$ the cell's mass density, $V_i$ the cell's volume, $B_{\lambda}$ is the Planck function, $T_i$ the cell's temperature, and $c$ is the speed of light. The opacity $\kappa_{i}$ has units cm$^{2}$/g. Note that the line-of-sight velocity $v_{\text{los}}$ is different for each cell at \emph{each} orbital phase because of the change in viewing geometry as the planet moves along its orbit\footnote{The derivation of the formula for $v_{\text{los}}$ can be found in Appendix B in \citet{Wardenier2021}. All spectra are calculated in the planet rest frame, without Doppler shifts due to orbital motion.}.    

In traditional Monte-Carlo radiative transfer, the number of photon packets $N_{i}$ that are ``sourced'' from a cell is proportional to the cell's luminosity, i.e., $N_{i} \propto L_i$. However, in the context of UHJs, this approach is not ideal as the large temperature contrast would cause the nightside to be undersampled relative to the dayside, leading to stochastic noise in the modeled spectrum. To circumvent this issue, gCMCRT uses ``composite biasing'' (\citealt{Baes2016,Lee2017}), such that the number of photon packets per cell becomes

\begin{equation} \label{eq:bias}
N_i = N_{\text{tot}} \bigg( (1-\xi) \frac{L_i}{L_{\text{tot}}} + \xi \frac{1}{n_{\text{cells}}} \bigg).
\end{equation}

\noindent Here, $\xi \in [0,1]$ is a scaling parameter and $n_{\text{cells}}$ is the total number of atmospheric cells in the model. Furthermore, $L_{\text{tot}} = \Sigma L_{i}$ is the total luminosity of the atmosphere and $N_{{\text{tot}}} = \Sigma N_{i}$ is the total number of emitted photon packets at the given wavelength ($N_{{\text{tot}}} = 10^6$ in this work). Equation \ref{eq:bias} is a compromise between non-biased Monte-Carlo radiative transfer \mbox{($\xi = 0$)} and uniform sampling \mbox{($\xi = 1$)}, where each cell emits the same number of packets. In this work, we use \mbox{$\xi = 0.99$}. To ensure that the total luminosity of the atmosphere is conserved, each packet is assigned a weight $w_i$ calculated as

\begin{equation} \label{eq:weights}
w_i = \frac{1}{(1-\xi)+\xi \langle L_i \rangle/L_i} = \bigg( \frac{N_{\text{tot}}}{N_{i}} \bigg) \bigg( \frac{L_{i}}{L_{\text{tot}}} \bigg), 
\end{equation}

\noindent with $\langle L_i \rangle$ the average luminosity of all cells. The second equality holds for any value of $\xi$.

\begin{figure*}
\vspace{-8pt} 

\makebox[\textwidth][c]{\hspace{-6pt} \includegraphics[width=1.01\textwidth]{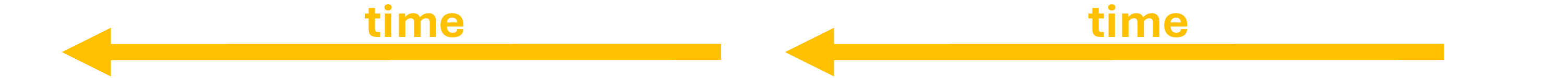}}

\vspace{-31pt} 

\makebox[\textwidth][c]{\hspace{-18pt}\includegraphics[width=1.23\textwidth]{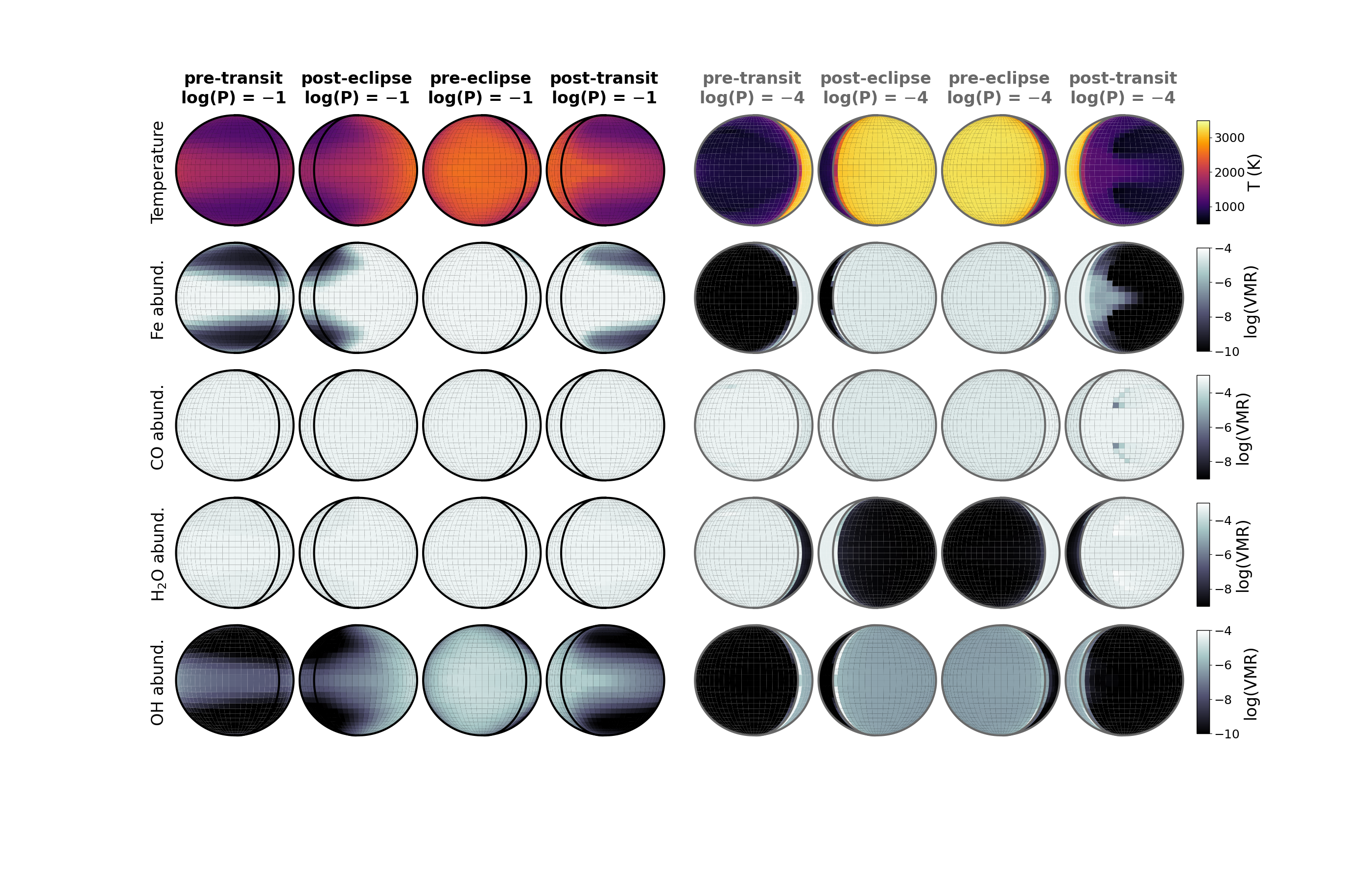}}

\vspace{-67pt}

\caption{2D plots of the atmospheric structure of the drag-free WASP-76b model from the SPARC/MITgcm. The top row shows the temperature structure, while the other rows show the abundances of Fe, CO, H$_2$O, and OH, respectively. Each column corresponds to a combination of orbital phase (pre-transit, post-transit, pre-eclipse, or post-eclipse) and pressure. The four columns on the left pertain to \mbox{$P$ = 10$^{-1}$} bar and the four columns on the right to $P$ = 10$^{-4}$ bar. The time arrow points from right to left to ensure consistency with Fig. \ref{fig:schematic_orbit}, in which the planet moves from right to left from the observer's perspective. During pre-transit \mbox{($\phi$ = 315$^\circ$)} and post-transit \mbox{($\phi$ = 45$^\circ$)}, the majority of the cooler nightside is in view. Most of the hotter dayside is visible during pre-eclipse \mbox{($\phi$ = 135$^\circ$)} and post-eclipse ($\phi$ = 225$^\circ$). In each plot, the solid arch marks the terminator between the dayside and the nightside of the planet.} 
\label{fig:globes_abunds}
\end{figure*}

Once a photon packet is sourced from a random position in cell $i$, gCMCRT evaluates the total optical depth $\tau$ encountered by the packet as it escapes from the atmosphere along the line of sight. To account for the effects of winds and planet rotation, we Doppler-shift the opacities in each atmospheric cell traversed by the photon packet according to the local line-of-sight velocity. This calculation involves the same equations as those presented in Section 3.3 in \citet{Wardenier2021}. The resulting ``peel-off'' weight $w_{\text{po},i}$ of the photon packet as it exits the atmosphere is (e.g., \citealt{Lee2017})

\begin{equation}
w_{\text{po},i} = \frac{1}{4\pi} w_i e^{-\tau} =  \bigg( \frac{N_{\text{tot}}}{N_i} \bigg) \bigg( \frac{ \rho_i V_i \kappa_i B_{\lambda}(T) }{L_{\text{tot}}} \bigg) e^{-\tau},
\end{equation}

\noindent with $1/4\pi$ a factor accounting for isotropic emission. Finally, the flux escaping from the planet disk (with units erg/cm$^2$/s/cm) into the observational direction is obtained by performing a sum over the photon packets emerging from all cells:

\begin{equation} \label{emission_flux}
F_{\lambda} = \bigg( \frac{L_{\text{tot}}}{R_{\text{p}}^2}  \bigg) \bigg( \sum_j^{N_{\text{tot}}} w_{\text{po},j} \bigg), 
\end{equation}

\noindent with $R_{\text{p}}$ the planet radius corresponding to the top of the atmosphere. To make our computation more efficient, we do not source photon packets from cells that are more than $\tau = 30$ away from the top of the atmosphere along the line of sight. These cells are either situated in very deep layers of the atmosphere or on the ``far side'' of the planet that is not in view. 

\subsection{Wavelength range, resolution, and opacities}

We simulate the thermal emission spectra of our models across the ESPRESSO (0.38--0.79 $\mu$m) and IGRINS (1.43--2.42 $\mu$m) bandpasses. These are the same wavelength ranges that were used by \citet{Wardenier2023} to compute \emph{transmission} spectra of \mbox{WASP-76b}. The optical spectra are simulated at a spectral resolution \mbox{$R$ = 300,000} ($>$$2\times$ the ESPRESSO resolution) and the infrared spectra at \mbox{$R$ = 135,000} ($\sim$3$\times$ the IGRINS resolution). Additionally, we compute a few individual spectral lines at \mbox{$R$ = 500,000} to resolve their shapes in more detail (see Figs. \ref{fig:emis_lines_Fe} to \ref{fig:emis_lines_OH}). 

In both bandpasses, we also use the same opacities as \citet{Wardenier2023}. These include continuum opacities due to H$_2$, He, and H scattering, collision-induced absorption (CIA) by H$_2$-H$_2$ and H$_2$-He, and bound-free and free-free transitions of H$^-$ (for references, see Table 2 in \citealt{Lee2022b}), as well as line absorption from Fe, Fe \textsc{ii}, K, Na, Ti, Mn, Mg, Cr, Ca \textsc{ii} (all from \citealt{Kurucz1995}), TiO (\citealt{Mckemmish2019}), VO (\citealt{McKemmish2016}), CO (\citealt{Li2015}), H$_2$O (\citealt{Polyansky2018}), OH (\citealt{Rothman2010}), CH$_4$ (\citealt{Hargreaves2020}), CO$_2$ (\citealt{Rothman2010}), HCN (\citealt{Barber2014}), and NH$_3$ (\citealt{Coles2019}). For further details about the opacities and the radiative transfer, we refer to Section 2.2 in \citet{Wardenier2023}.

\begin{figure*} 
\makebox[\textwidth][c]{\hspace{-12pt}\includegraphics[width=1.2\textwidth]{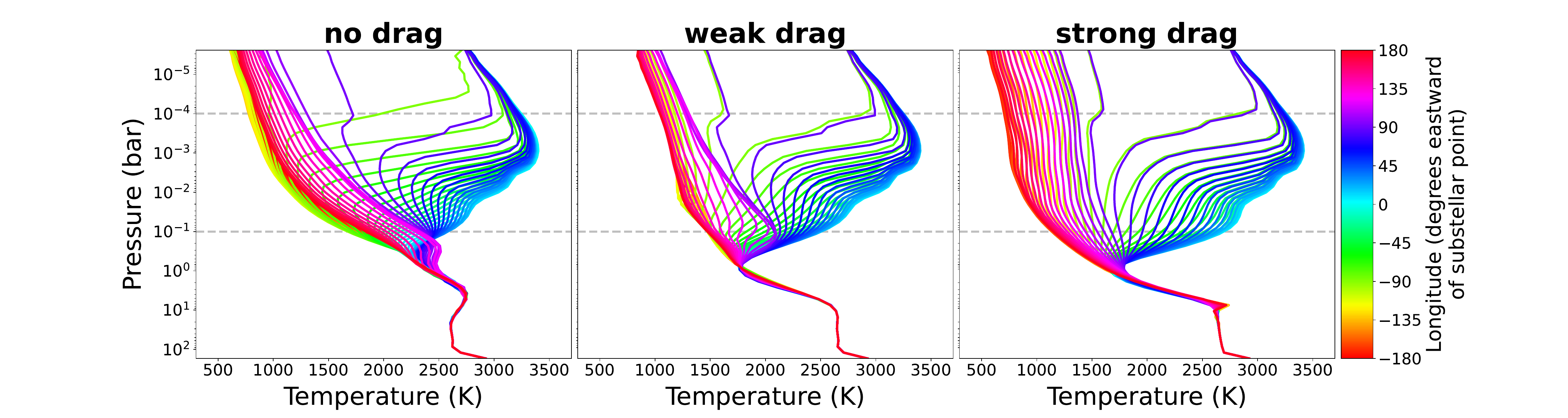}}

\vspace{-3pt}

\caption{Equatorial temperature profiles for each of the three WASP-76b models. The color scale indicates the longitude (eastward of the substellar point) to which a profile corresponds. Longitudes with a negative value lie on the western hemisphere, while longitudes with a positive value lie on the eastern hemisphere. In the drag-free model, the temperature profile on the western dayside (in lime green) has a much stronger vertical gradient compared to the eastern dayside (in dark blue). The horizontal dashed lines correspond to pressures \mbox{$P$ = 10$^{-1}$} bar and $P$ = 10$^{-4}$ bar, for which the atmospheric structure is plotted in Figs. \ref{fig:globes_abunds} and \ref{fig:globes_vlos} .}  
\label{fig:equatorial_PT}
\end{figure*}

\subsection{Cross-correlation and K$_{\text{p}}$--V$_{\text{sys}}$ maps}
\label{subsec:CCF_description}

Once we obtained all spectra $\vec{x}(\phi)$ as a function of orbital phase, we compute the 2D cross-correlation function (CCF; \citealt{Snellen2010}) by multiplying the spectra with a template $\vec{T}$ shifted by different velocities $v$:

\begin{equation} \label{eq:def_cross_correlation}
\text{CCF}(\phi, v) = \sum_{j}^{N_\lambda} x_j(\phi) \ T_j(v),
\end{equation}

\noindent where $N_\lambda$ is the number of wavelength points. To compute the template of a species $X \in \{$Fe, CO, H$_2$O, OH$\}$, we only include the line absorption from $X$ plus continuum opacities in the radiative transfer.

For each species, we generate two templates, which are derived from the drag-free model: the static \emph{dayside} spectrum at $\phi = 180^\circ$ and the static \emph{nightside} spectrum at $\phi = 0^\circ$. To obtain these ``static" spectra, we set the line-of-sight velocities to zero in the radiative transfer (this means that we also remove the effect of planet rotation from the templates). Because the templates do not depend on orbital phase, they do not account for the fact that different parts of the 3D temperature structure are in view at different phases. Hence, the templates are essentially one-dimensional. Before computing the CCF, we subtract the continuum from both the spectra and the templates by convolving them with a Gaussian kernel and subtracting the result. After this, we ``standardize'' the individual templates by subtracting the mean and dividing by the standard deviation. We evaluate the CCF for $v \in [-60,60]$ km/s, with steps of 0.25 km/s.

For each CCF map, we calculate four $K_{\text{p}}$--$V_{\text{sys}}$ maps associated with the pre-transit \mbox{($285^\circ < \phi < 345^\circ$)}, post-transit ($15^\circ < \phi < 75^\circ$), pre-eclipse ($105^\circ < \phi < 165^\circ$), and post-eclipse ($195^\circ < \phi < 255^\circ$), respectively. To this end, we co-add the CCF values along orbital trails of the form $v(\phi) = \Delta V_{\text{sys}} + \Delta K_{\text{p}}\sin(\phi)$ between the specified orbital-phase angles (see also \citealt{Wardenier2023}). This gives

\begin{equation} \label{eq:kpvsys_sum}
    \text{SNR}(\Delta K_{\text{p}}, \Delta V_{\text{sys}}) = \frac{1}{\alpha} \sum_i^{N_\phi} \text{CCF} \Big(\phi_i, v(\phi_i) \Big), 
\end{equation}

\noindent where SNR is the value of the $K_{\text{p}}$--$V_{\text{sys}}$ map at the point $(\Delta K_{\text{p}},\Delta V_{\text{sys}})$, {\color{black}{$\alpha$ is a scaling factor such that the maximum is 1}}, and $N_{\phi}$ is the number of emission spectra over which the sum is performed. As shown in Fig. \ref{fig:schematic_orbit}, $N_{\phi} = 6$ for each of the orbital phases that we consider. At each phase angle, we obtain the CCF value at $v(\phi_i)$ by interpolating between the two values at the nearest velocity shifts in the CCF map.

\begin{figure*}[t!]

\vspace{-8pt} 

\makebox[\textwidth][c]{\hspace{-6pt} \includegraphics[width=1.01\textwidth]{figures/timearrow.pdf}}

\vspace{-21pt} 

\makebox[\textwidth][c]{\hspace{-18pt}\includegraphics[width=1.23\textwidth]{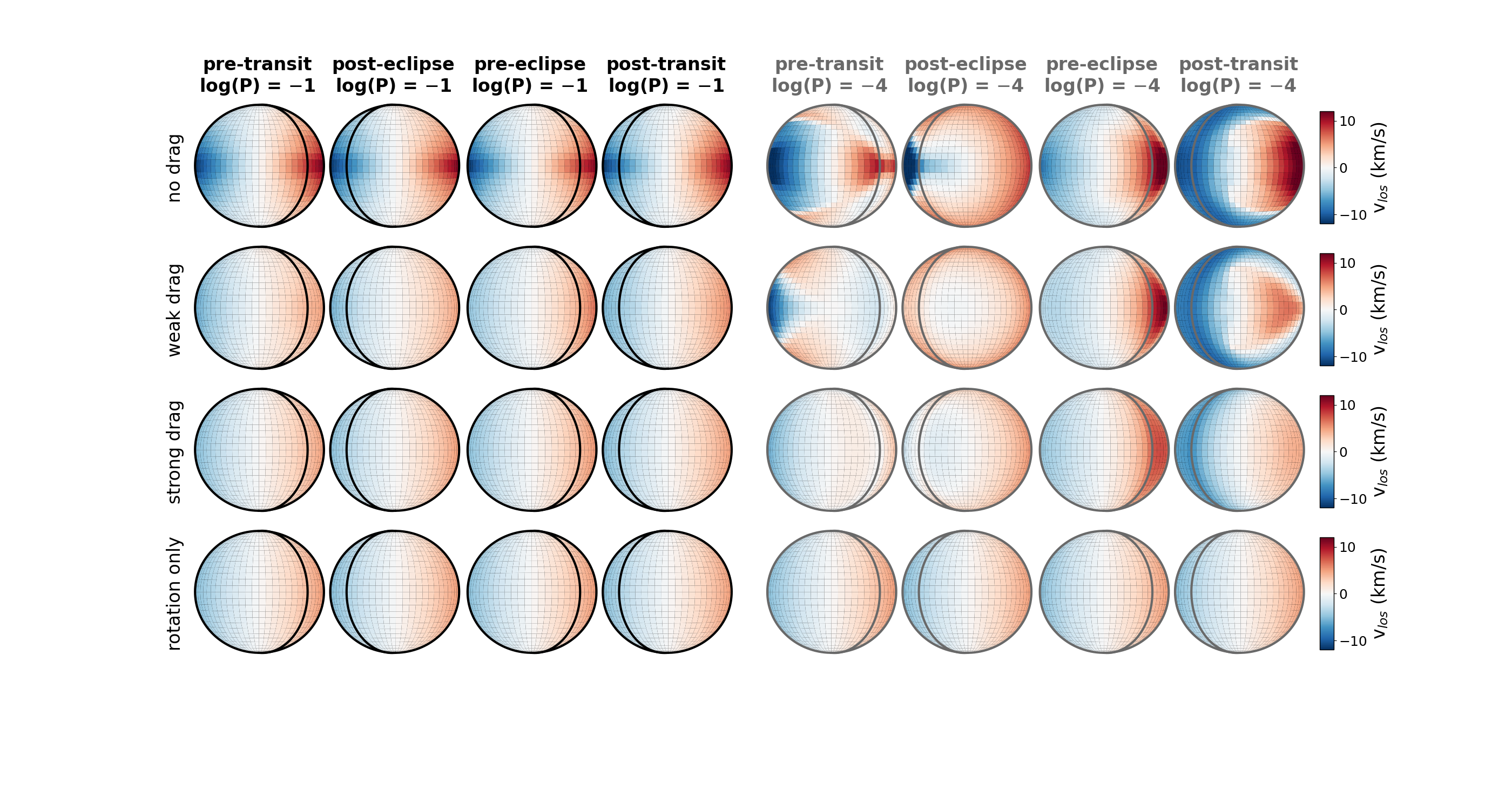}}

\vspace{-60pt}

\caption{Same as Fig. \ref{fig:globes_abunds}, but now showing the line-of-sight velocities $v_{\text{los}}$ due to winds and rotation. The top three columns correspond to the drag-free model, the weak-drag model, and the strong-drag model of WASP-76b, respectively. The bottom row shows the line-of sight velocities across the planet disk when only considering the contribution from rotation. A blueshift ($v_{\text{los}} < 0$) implies that the atmosphere is moving towards the observer, while a redshift ($v_{\text{los}} > 0$) is induced by material moving away along the line of sight.}  
\label{fig:globes_vlos}
\end{figure*}

\section{Results \& Discussion}
\label{sec:results}

\subsection{Global circulation models}

Fig. \ref{fig:globes_abunds} shows the temperature structure and the abundance distributions of the three WASP-76b models. The plots depict the 2D planet disk at four different orbital phases and two different pressures ($10^{-1}$ and $10^{-4}$ bar). In the IGRINS bandpass, the emission spectrum will typically probe pressures as high as $\sim$$10^{-1}$ bar (e.g., Fig. 2 in \citealt{Smith2024}). In the optical, Fe lines can probe lower pressures in the millibar regime (e.g., Fig. 4 in \citealt{Pino2020}). {\color{black}{In Appendix \ref{app:A}, we present individual contribution functions for the chemical species considered in this work.}} Fig. \ref{fig:equatorial_PT} shows the 1D temperature profiles of the three models in the equatorial plane. As illustrated in both Figs. \ref{fig:globes_abunds} and \ref{fig:equatorial_PT}, the day-night temperature contrast of the atmosphere increases sharply with altitude. At $10^{-1}$ bar, the dayside temperature is still low enough for H$_2$O to persist. However, at $10^{-4}$ bar, the dayside is completely depleted of H$_2$O due to thermal dissociation. On the other hand, the nightside temperature \emph{decreases} with altitude, resulting in Fe being condensed out of the atmosphere at $10^{-4}$ bar. Owing to its strong triple bond, CO is neither prone to condensation nor dissociation, leading to a uniform 3D abundance (e.g., \citealt{Beltz2022,Savel2022}). Consequently, any changes in the line strength of CO should directly probe the planet's 3D temperature structure. 

Finally, Fig. \ref{fig:equatorial_PT} shows that the nightside temperature profile becomes more isothermal with increasing drag strength. {\color{black}{This can be understood intuitively. Stronger drag leads to less efficient heat redistribution, and thus less energy being deposited onto the nightside. Furthermore, the radiative timescale $\tau_\text{rad}$ (the time it takes for a parcel of air to lose a substantial fraction of its energy by radiation) is proportional to $T_\text{photo}^{-3}$, with $T_\text{photo}$ the photospheric temperature (\citealt{PerezBecker2013,Parmentier2021}). Because the nightside photosphere is cooler and radiating less energy into space per unit time, radiative equilibrium will converge to a shallower vertical temperature gradient.}} As we will demonstrate in the next sections, the strength of the vertical temperature gradient impacts the ``detectability'' of the nightside spectrum via high-resolution observations. 

Fig. \ref{fig:globes_vlos} shows the line-of-sight velocities in each model due to planet rotation and the 3D wind profile. In the drag-free model (top row), the equatorial jet is clearly visible at both pressures. $v_{\text{los}}$ can reach values of $\pm$10 km/s. As the drag strength increases, the wind speeds \emph{decrease} and the line-of-sight velocities become dominated by solid-body rotation (bottom row). The equatorial rotation velocity of WASP-76b is about $\pm$5.3 km/s.

\begin{figure*}
\makebox[\textwidth][c]{\hspace{-12pt}\includegraphics[width=1.0\textwidth]{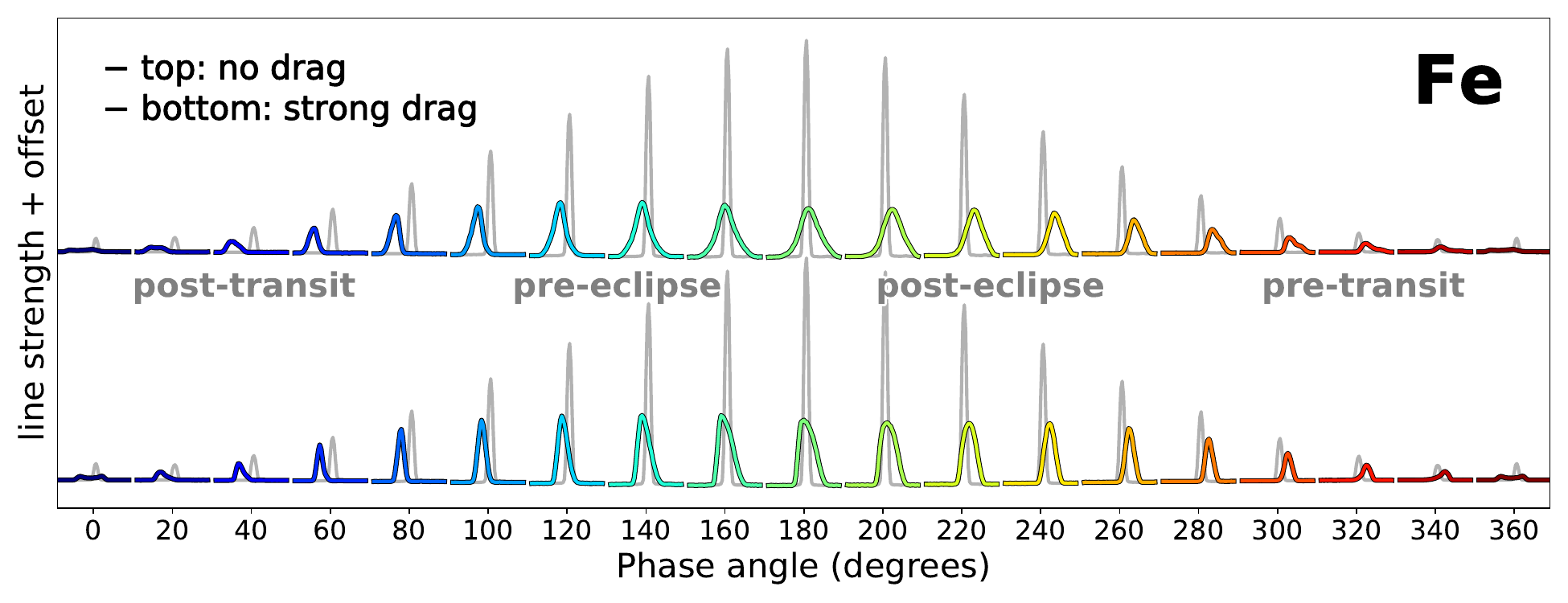}}
\vspace{-18pt}
\caption{The shape of a \emph{single} Fe line (0.54751 -- 0.54757 $\mu$m) in the rest frame of WASP-76b as a function of orbital phase (computed at $R$ = 500,000). The Fe lines in the top row were obtained from the drag-free model, while the Fe lines in the bottom row are from the model with strong drag. The colored lines include the effects of Doppler shifts due to winds and rotation. On the other hand, the grey lines in the background correspond to a ``static" atmosphere in which the line-of-sight velocities are zero. Because the colored lines are subject to Doppler effects, they appear broadened, shifted, and distorted compared to the static spectrum. The transit and eclipse occur at phase angles \mbox{$\phi$ = 0$^\circ$/360$^\circ$} and \mbox{$\phi$ = 180$^\circ$}, respectively.} 
\label{fig:emis_lines_Fe}
\end{figure*}

\begin{figure*}
\makebox[\textwidth][c]
{\hspace{-12pt}\includegraphics[width=1.0\textwidth]{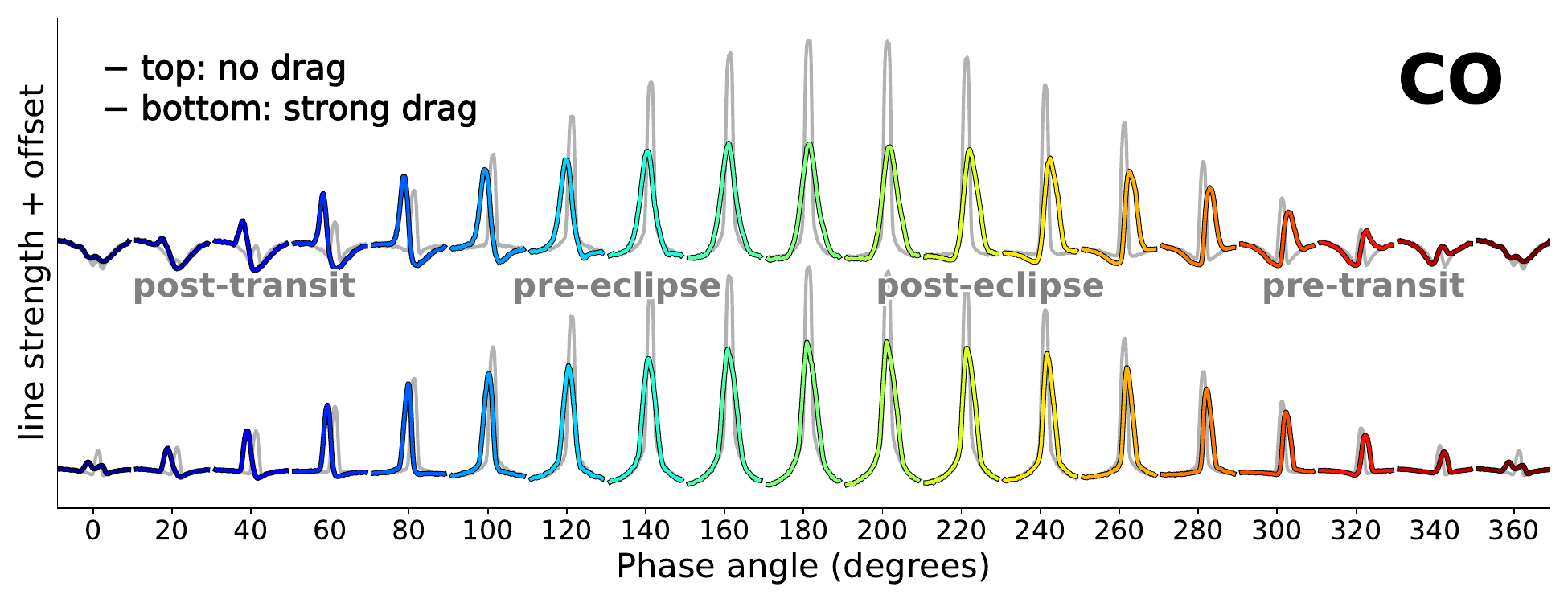}}
\vspace{-18pt}
\caption{Same as Fig. \ref{fig:emis_lines_Fe}, but now for a \emph{single} CO line (2.3019 -- 2.3023 $\mu$m) in the emission spectrum of WASP-76b.}  
\label{fig:emis_lines_CO}
\end{figure*}

\subsection{Individual spectral lines}
\label{subsec:spectral_lines}

\begin{figure*}
\makebox[\textwidth][c]{\hspace{-12pt}\includegraphics[width=1.0\textwidth]{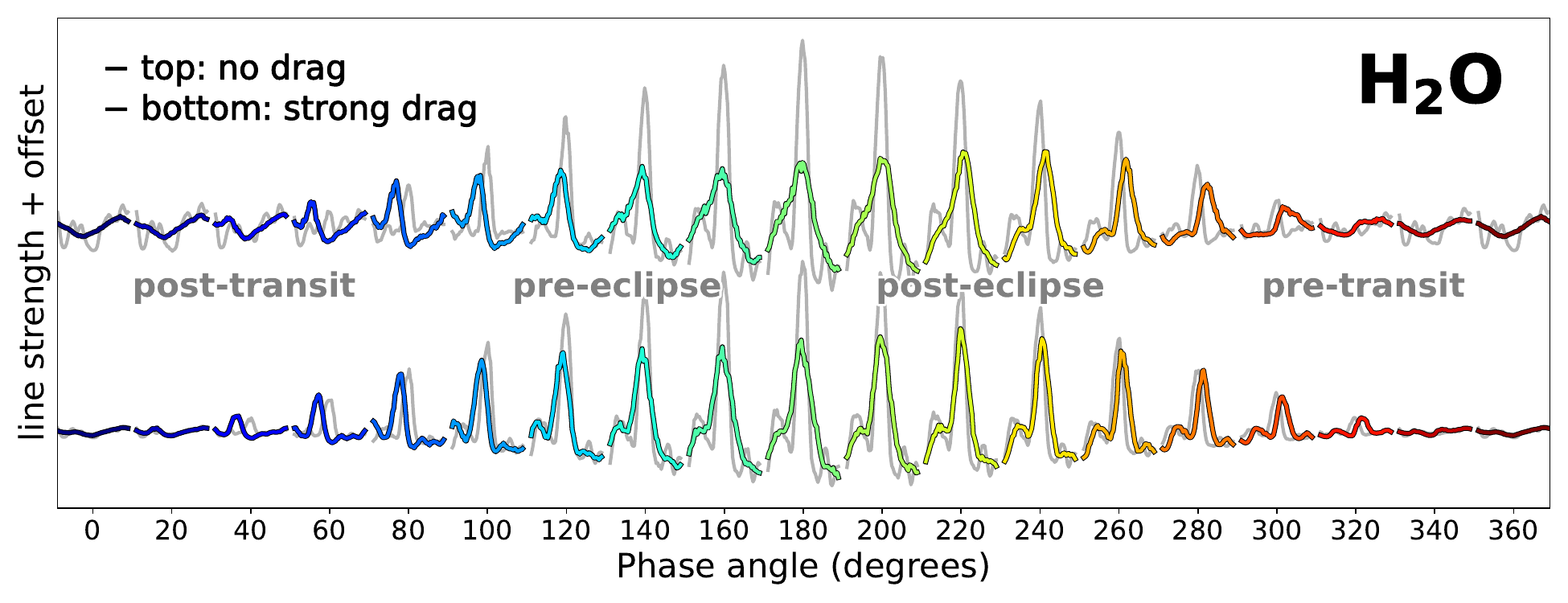}}
\vspace{-18pt}
\caption{Same as Figs. \ref{fig:emis_lines_Fe} and \ref{fig:emis_lines_CO}, but now for a \emph{single} H$_2$O line (1.7831 -- 1.7833 $\mu$m) in the emission spectrum of WASP-76b.}  
\label{fig:emis_lines_H2O}
\end{figure*}

\begin{figure*}
\makebox[\textwidth][c]{\hspace{-12pt}\includegraphics[width=1.0\textwidth]{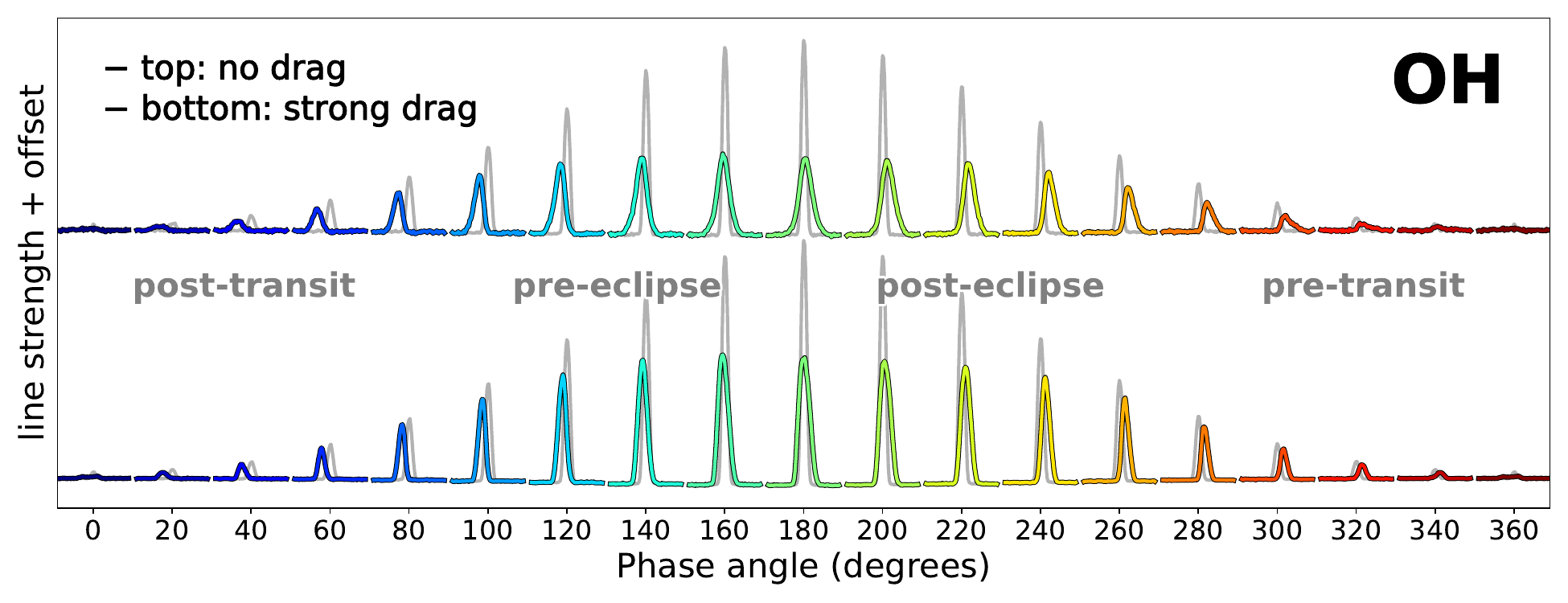}}
\vspace{-18pt}
\caption{Same as Figs. \ref{fig:emis_lines_Fe}, \ref{fig:emis_lines_CO}, and \ref{fig:emis_lines_H2O}, but now for a \emph{single} OH line (1.80183 -- 1.8021 $\mu$m) in the emission spectrum of WASP-76b.}  
\label{fig:emis_lines_OH}
\end{figure*}

Figs. \ref{fig:emis_lines_Fe} to \ref{fig:emis_lines_OH} show individual spectral lines of Fe, CO, H$_2$O, and OH as a function of orbital phase in the rest frame of WASP-76b. Each plot depicts the lines associated with the drag-free model (top row) and the strong-drag model (bottom row), respectively. The spectral lines of the static atmospheres (with $v_{\text{los}}$ = 0 km/s) are shown in the background for reference. The first thing to note is that the lines accounting for Doppler effects have a lower line strength compared to their static counterparts. This is a result of Doppler broadening, which is mostly driven by planet rotation. Also, because the line-of-sight velocities in the drag-free model have a larger dispersion due to the wind profile (see Fig. \ref{fig:globes_vlos}), the lines in the spectrum of the drag-free model are slightly broader compared to the strong-drag model. As expected, the emission lines of all species are strongest around the eclipse, when the full dayside of the planet is in view. 

Fig. \ref{fig:emis_lines_Fe} depicts a single Fe line. At every orbital phase, it appears in emission (that is, the flux from the line core is larger than the flux  from the continuum), demonstrating that Fe exclusively probes the thermally inverted dayside of the planet. This result is compatible with Fe condensation on the nightside (see Fig. \ref{fig:globes_abunds}). During post-transit and pre-eclipse, the emission lines are blueshifted in the planet rest frame as the dayside rotates \emph{towards} the observer. During post-eclipse and pre-transit, the dayside rotates \emph{away} from the observer, inducing a redshift. Remarkably, we can still see weak emission lines at orbital phases $\pm$$20^\circ$, which must originate from a very thin slice of dayside atmosphere that is still in view.

Fig. \ref{fig:emis_lines_CO} shows an example of a CO line. Contrary to Fe, the abundance of CO is uniform across the atmosphere, meaning that the CO signal should probe both the dayside and the nightside. During pre- and post-eclipse, the CO lines appear purely in emission, as the strongest signal emerges from the dayside. During pre- and post-transit, however, the spectral-line shapes are more intricate. In the drag-free model, the lines consist of a component in absorption (associated with the nightside) and a component in emission (associated with the dayside). The components exhibit opposite Doppler shifts as the dayside and the nightside have different line-of-sight velocities due to planet rotation. Furthermore, the CO lines in pre- and post-transit appear as each other's ``mirrored'' versions, as the dayside is redshifted before the transit, but blueshifted after the transit (and vice versa for the nightside). At phases $\pm$$40^\circ$, the absorption and emission components are roughly equally strong, even though a much larger fraction of the nightside is in view. Closer to the transit, the lines appear mainly in absorption. However, the absorption features are substantially weaker than the emission lines around the eclipse. This is a result of the lower nightside temperature and weaker vertical temperature gradient. A more extreme case is the strong-drag model, in which the absorption features of CO fully disappear. This is because the strong-drag model has a more isothermal temperature profile on its nightside than the drag-free model (see Fig. \ref{fig:equatorial_PT}). Therefore, in reality, the detectability of a planet's nightside absorption features via high-resolution spectroscopy (e.g., \citealt{Mraz2024,Yang2024}) will very much depend on the slope of the nightside temperature profile, set by the efficiency of heat redistribution across the atmosphere.

\begin{figure*}
\vspace{0pt}
{\hspace{0pt}\includegraphics[width=1.01\textwidth]{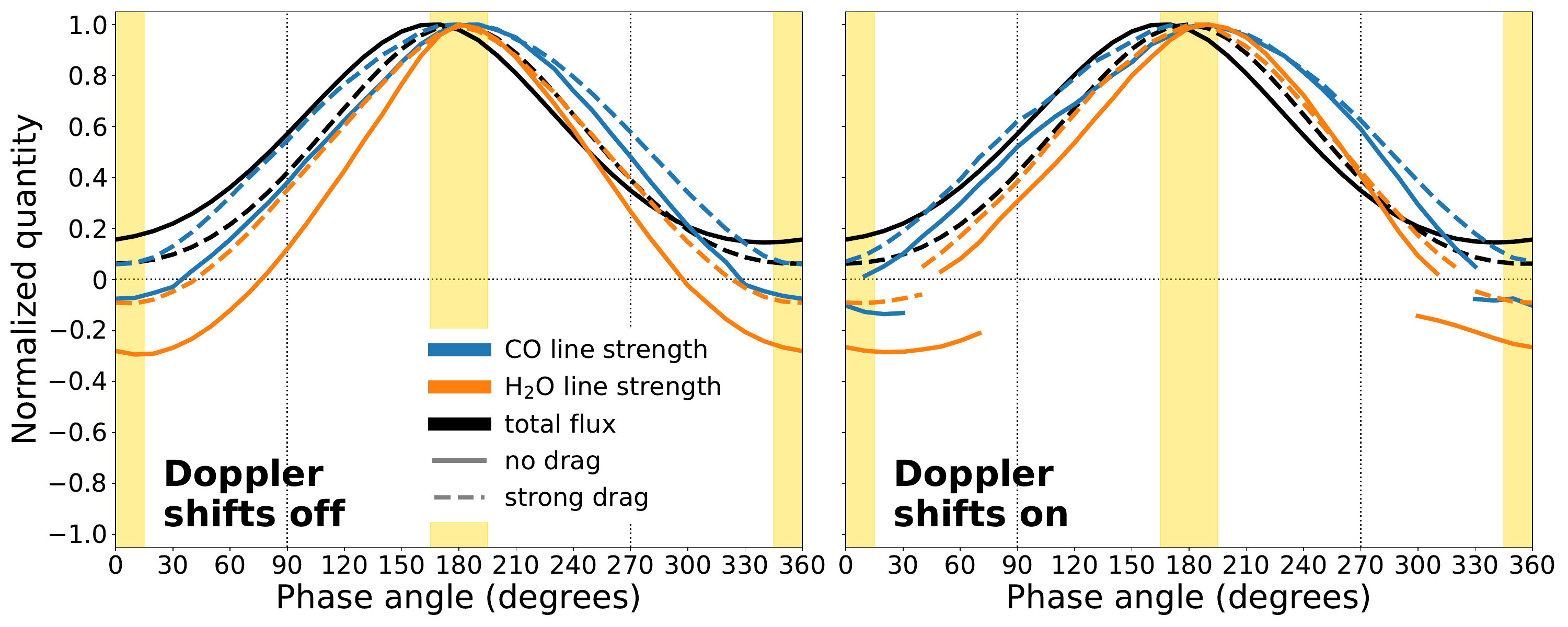}}
\vspace{-15pt}
\caption{Phase-dependent line strengths of CO and H$_2$O in the drag-free model (solid curves) and the strong-drag model (dashed curves), respectively. The left panel shows the line strengths for the static model spectra \mbox{($v_{\text{los}}$ = 0 km/s)}, while the right panel shows the line strengths when accounting for Doppler shifts. A positive line strength implies that the spectrum contains \emph{emission} lines, while a negative line strength implies \emph{absorption} lines. In the right panel (especially in post-transit; $\phi < 90^\circ$), there are phases at which emission features from the dayside and absorption features from the nightside coexist, as they are subject to different Doppler shifts. The (wavelength-integrated) phase curves of both models across the IGRINS bandpass are plotted in black. The yellow regions depict phases during which the emission spectrum is not observable due to the transit and secondary eclipse. Before calculating the line strengths shown in this figure, we convolved the spectra and the cross-correlation templates with a Gaussian kernel corresponding to the IGRINS resolution ($R$ = 45,000).}
\label{fig:line_contrast_curves}
\end{figure*}

\begin{figure}
\vspace{-4pt}
{\hspace{-11pt} \includegraphics[width=0.54\textwidth]{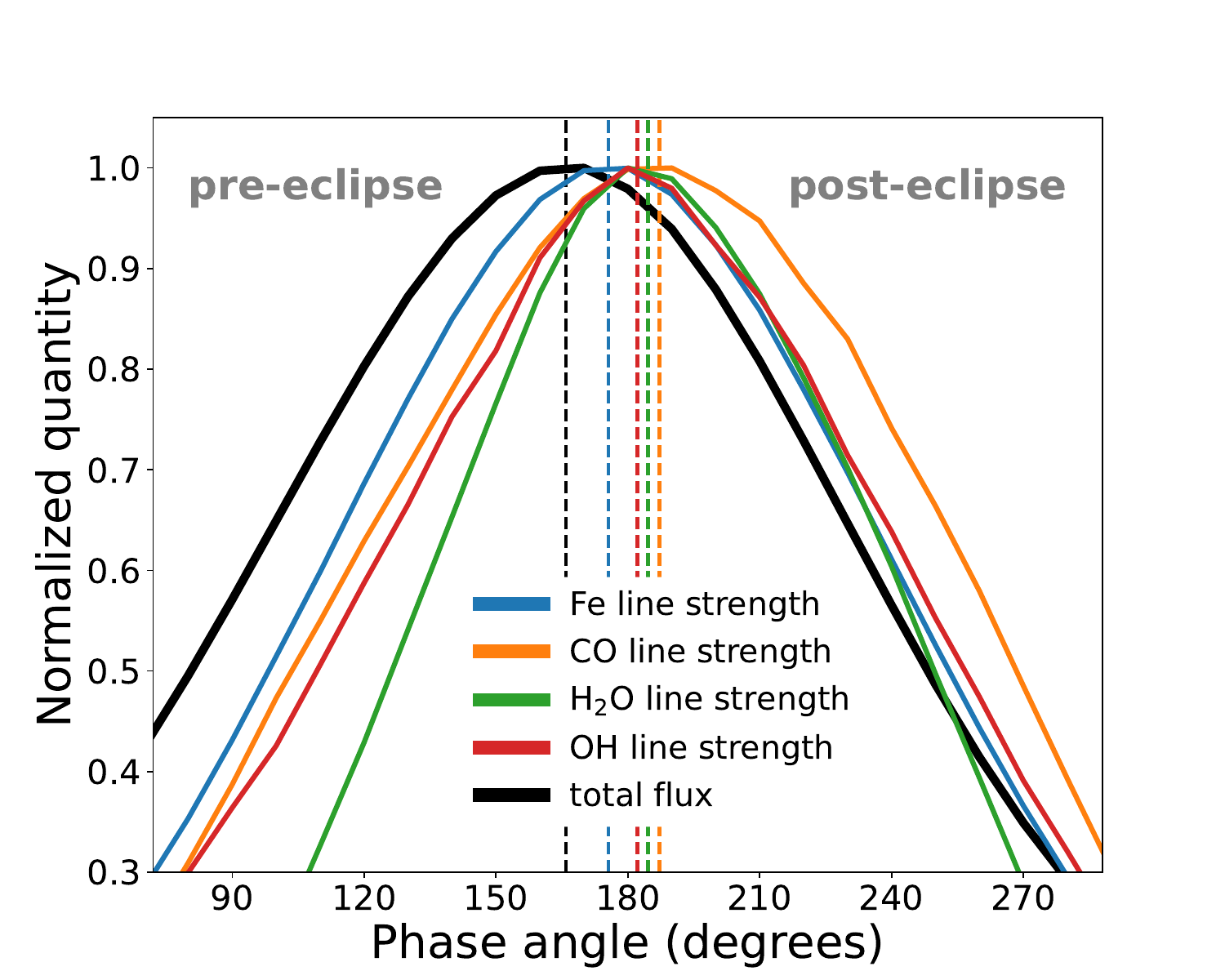}}
\vspace{-15pt}
\caption{The phase curve of the drag-free WASP-76b model (in black), with the phase-dependent line strengths of Fe, CO, H$_2$O, and OH superimposed (in color, assuming no Doppler shifts). While the total emitted flux and the Fe line strength peak \emph{before} the eclipse ($\phi < 180^\circ$), the emission lines of CO, H$_2$O and OH are strongest \emph{after} the eclipse ($\phi > 180^\circ$). The dashed lines indicate the maximum of each curve and were obtained from Gaussian fits.}
\label{fig:phase_curve}
\end{figure}

Fig. \ref{fig:emis_lines_H2O} shows a single H$_2$O line. While H$_2$O is present on both the dayside and the nightside of the planet, it is dissociated at lower pressures on the dayside due to the thermal inversion. Again, we can see that H$_2$O appears in emission during pre- and post-eclipse. As with the CO lines, the strongest differences between the drag-free and the strong-drag model occur during pre- and post-transit. In the drag-free model, the (``V-shaped'') absorption features associated with the nightside are clearly visible. However, in the strong-drag model, the line strength of the absorption component of the H$_2$O lines is very marginal. Even though H$_2$O is abundant on the nightside, the shallow gradient of the temperature profile can wipe out its spectral features as the black-body temperatures probed by the line core and the continuum are very similar. Unlike Fe and CO, H$_2$O does not show up in emission at $\pm$$20^\circ$ in the strong-drag model. This is likely an effect of thermal dissociation: H$_2$O lines probe deeper layers on the dayside that are geometrically masked by the nightside at these orbital phases.

\begin{figure*}
\vspace{-30pt} 
\makebox[\textwidth][c]{\hspace{-1pt}\includegraphics[width=1.15\textwidth]{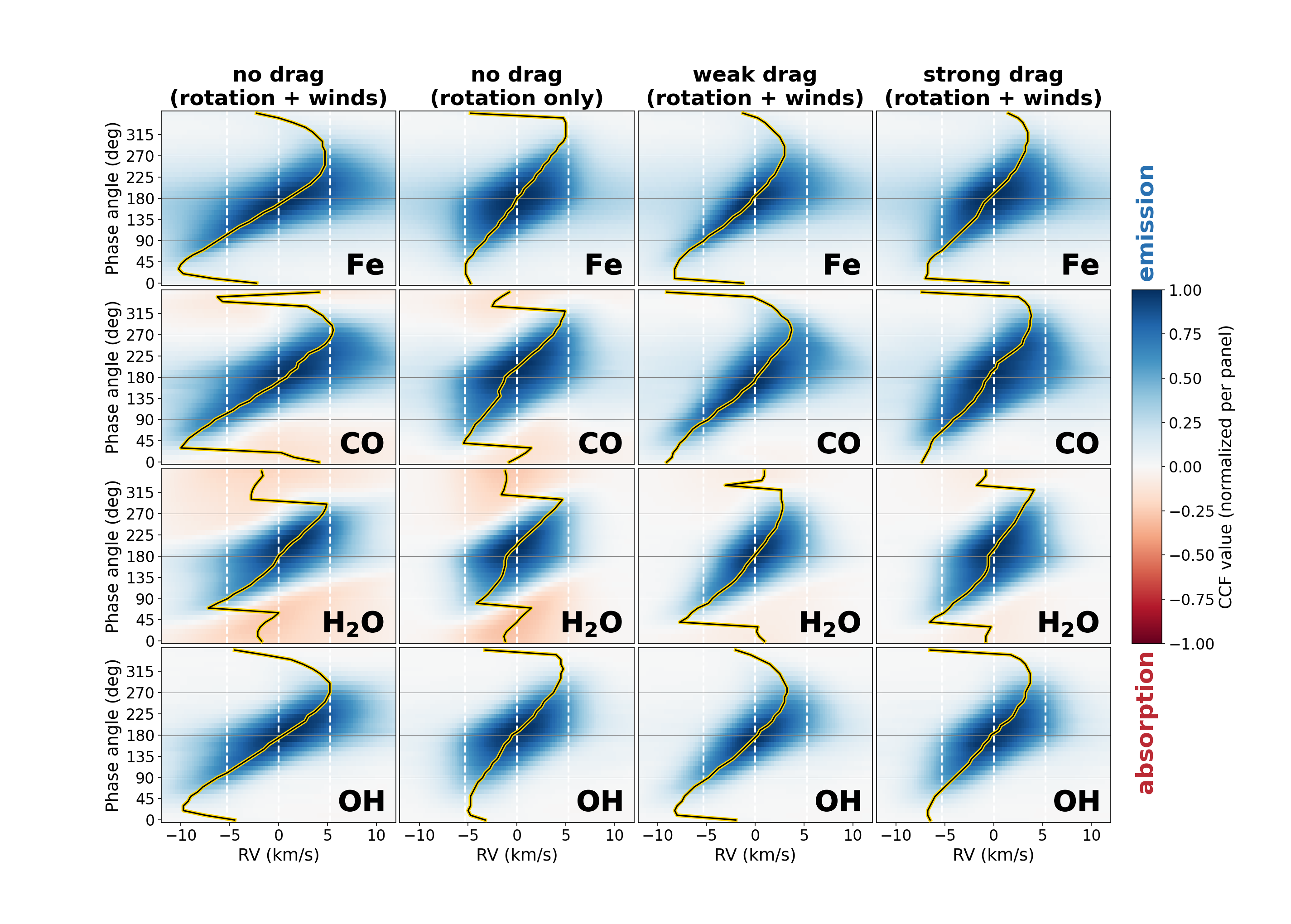}}
\vspace{-39pt}
\caption{CCF maps for Fe, CO, H$_2$O, and OH (rows) in the planet rest frame. We show the results for the three GCMs of WASP-76b, as well as the drag-free model with Doppler shifts due to rotation only (columns). Each panel shows the rest-frame CCF over the full orbit of the planet, where phase angles \mbox{$\phi$ = 0$^\circ$} and \mbox{$\phi$ = 180$^\circ$} correspond to the transit and the eclipse, respectively. To produce the CCF maps in this figure, \emph{all} spectra were cross-correlated with the static dayside emission templates from the drag-free model. In each map, a positive CCF value implies \emph{emission} features, while a negative CCF value implies \emph{absorption} features. The yellow trails mark the most extreme CCF values at each orbital phase. RV $<$ 0 corresponds to a net blueshift, while RV $>$ 0 corresponds to a net redshift. The vertical dashed lines are located at 0 km/s and $\pm$5.3 km/s, which is the rotational velocity at the equator.}  
\label{fig:ccf_maps_full_orbit}
\end{figure*}

Finally, Fig. \ref{fig:emis_lines_OH} shows an example of an OH line, whose behavior is very similar to that of the Fe line in Fig. \ref{fig:emis_lines_Fe}. It is worth mentioning that the line strength of the strong-drag model is not only higher due to the smaller effect of Doppler broadening, but also because it has a stronger dayside temperature inversion than the drag-free model (see Fig. \ref{fig:equatorial_PT}). 

\subsection{Phase dependence of the line strengths}
\label{subsec:line_strengths}

In this section, we study the (average) line strengths of different species as derived from their CCFs, rather than focusing on single spectral lines. The line strengths reported in Figs. \ref{fig:line_contrast_curves} and \ref{fig:phase_curve} are equal to the value of the CCF peak at a given orbital phase. All CCFs were obtained using the same dayside templates, which are described in Section \ref{subsec:CCF_description}.

\begin{figure*}
\vspace{-35pt} 
\makebox[\textwidth][c]{\hspace{30pt}\includegraphics[width=1.0\textwidth]{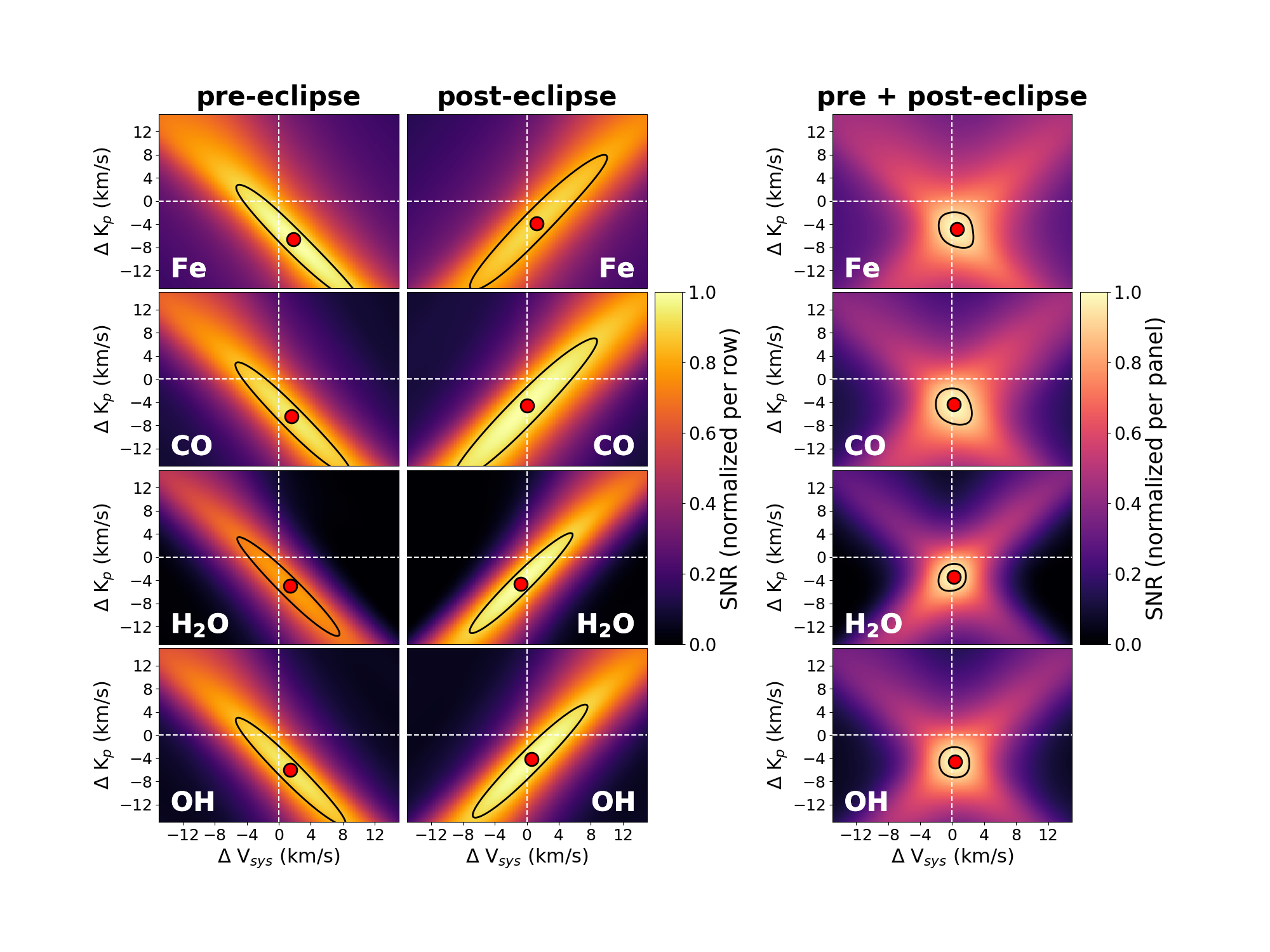}}
\vspace{-45pt}
\caption{$K_{\text{p}}$--$V_{\text{sys}}$ maps of the drag-free model of WASP-76b during the pre- and post-eclipse phases of the orbit (first two columns). The maps in the right column were obtained by adding the maps from the pre- and post-eclipse together. In each panel, the red marker indicates the ($\Delta$$K_{\text{p}}$, $\Delta$$V_{\text{sys}}$) position of the signal peak, while the black contour corresponds to 0.9$\times$ the peak value. In the absence of Doppler effects, the planet signal would be situated at \mbox{($\Delta$$K_{\text{p}}$, $\Delta$$V_{\text{sys}}$) =} (0, 0) km/s, indicated by the white dashed lines.}
\label{fig:kp_vsys_maps_dayside}
\end{figure*}

Fig. \ref{fig:line_contrast_curves} shows the phase-dependent line strengths of CO and H$_2$O in the drag-free and strong-drag models of WASP-76b, respectively. The left panel shows the line strengths when only accounting for the 3D temperature structure of the atmosphere, but not for Doppler shifts (similar to the ``static" lines in Figs. \ref{fig:emis_lines_Fe} to \ref{fig:emis_lines_OH}). In the drag-free model, H$_2$O lines appear in absorption during the majority of pre- and post-transit. For CO, the emission features are dominant over a much larger part of the orbit. In the strong-drag model, CO even appears in emission during the \emph{entire} orbit, which is in agreement with the (grey) spectral lines shown in the bottom row of Fig. \ref{fig:emis_lines_CO}. These results further underscore the challenges associated with detecting the nightside spectrum of UHJs. Not only are the absorption features constrained to smaller parts of the orbit (if the line strengths become negative at all), they are also weaker than the emission features visible during pre- and post-eclipse. Thirdly, the total thermal flux emanating from the nightside is at most $10$-$15\%$ of the dayside flux, resulting in lower photon counts. All things considered, detecting absorption features associated with UHJ nightsides may well be \emph{an order of magnitude harder} than detecting emission features from their daysides, especially since many UHJ observations are consistent with atmospheric drag (e.g., \citealt{Kreidberg2018,Arcangeli2019,Coulombe2023,Wardenier2024,Demangeon2024}). Therefore, their nightside temperature profiles can be expected to be relatively isothermal.

The right panel of Fig. \ref{fig:line_contrast_curves} shows the line strengths of CO and H$_2$O when also accounting for Doppler shifts in the radiative transfer. The main difference compared to the left panel is that there are now certain orbital phases at which the CCF has both a positive \emph{and} a negative peak (which occur at different effective Doppler shifts due to the rotation of the planet). This coexistence of absorption and emission features can also be seen in the spectral lines of CO and H$_2$O in Figs. \ref{fig:emis_lines_CO} and \ref{fig:emis_lines_H2O}.

Another takeaway from Fig. \ref{fig:line_contrast_curves} is that the line strengths are not perfectly correlated with the integrated flux from the planet. This is further illustrated in Fig. \ref{fig:phase_curve}, which shows the line strengths of the four species and the phase curve of the drag-free model in pre- and post-eclipse. To isolate the effect of the 3D temperature structure, we do not include Doppler shifts here. Because the planet has an eastward hotspot offset, the phase-curve maximum occurs \emph{before} the secondary eclipse. However, the line strengths of CO, H$_2$O, and OH are strongest \emph{after} the eclipse. The fact that all colored curves are offset from the black phase curve shows that the line strength is \emph{not} a measure of the absolute temperature probed by the spectrum. Rather, the line strength is set by the vertical temperature gradient of the atmosphere. This result confirms earlier findings by \citet{vanSluijs2022} based on models and observations of WASP-33b.

\begin{figure*}
\vspace{-10pt}
\makebox[\textwidth][c]
{\hspace{16pt}\includegraphics[width=1.19\textwidth]{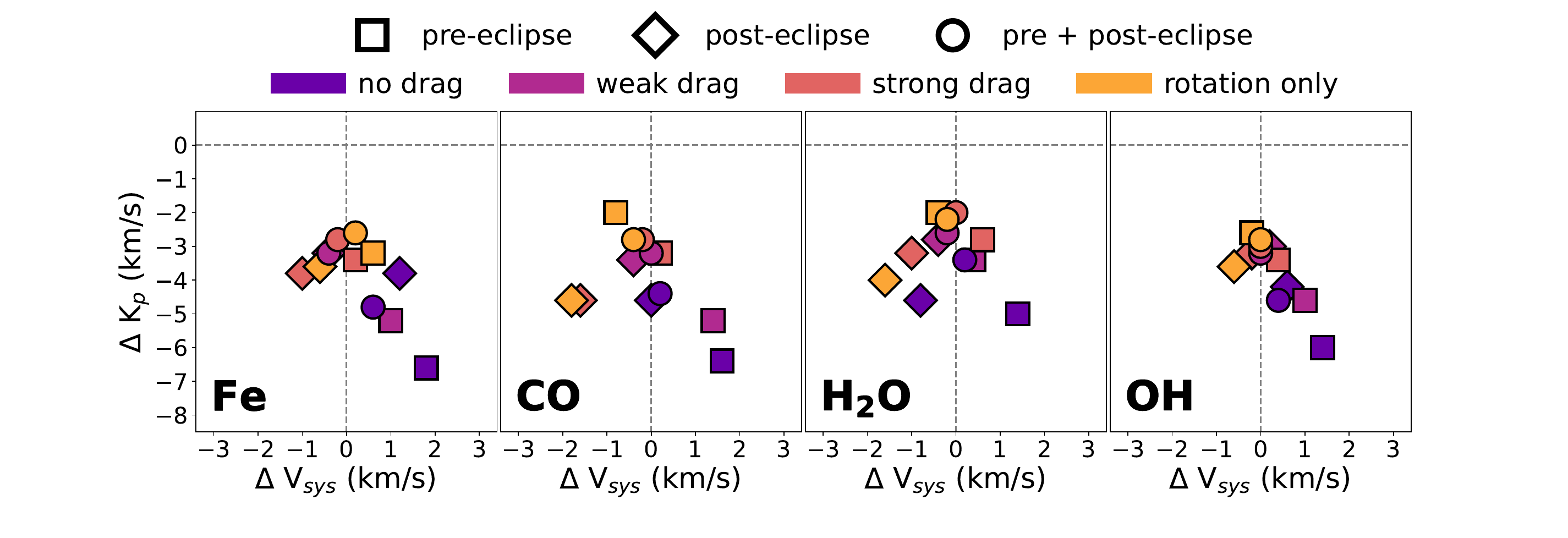}}
\vspace{-28pt}
\caption{Locations of the $K_{\text{p}}$--$V_{\text{sys}}$ maxima found for all species (one per panel) and all GCM outputs of WASP-76b (indicated by different colors). The symbol shapes indicate the orbital phases, while the symbol colors correspond to different models. The rotational velocity of WASP-76b at the equator is $\pm$5.3 km/s.}  
\label{fig:blob_distribution}
\end{figure*}

\subsection{CCF maps of the full orbit}

Fig. \ref{fig:ccf_maps_full_orbit} shows the CCF maps for all GCM outputs of WASP-76b. All maps were obtained by cross-correlating the model spectra with dayside emission templates from the drag-free model. For each of the four species, we recover the ``quasi-sinusoidal'' dependence of the net Doppler shifts on orbital phase (driven by planet rotation), as identified by \citet{Zhang2017}.

All CCFs show very similar behavior between the first and third quadratures of the orbit (i.e. $ 90^\circ \lesssim \phi \lesssim  270^\circ$). Around the first quadrature, all models show a strong blueshift (RV $<$ 0) in the planet rest frame. This is because all visible parts of the dayside rotate \emph{towards} the observer. As a larger part of the dayside comes into view, some regions will become redshifted \mbox{(see Fig. \ref{fig:globes_vlos})}, leading to a smaller net blueshift. As the planet approaches the eclipse, the average Doppler shift converges to zero. After the eclipse, the opposite effect occurs. At the third quadrature, the dayside is exclusively rotating \emph{away} from the observer, producing a strong redshift \mbox{(RV $>$ 0)}.

The second column of Fig. \ref{fig:ccf_maps_full_orbit} shows the CCFs when only accounting for Doppler shifts due to planet rotation. In this case, the net Doppler shifts never exceed the rotational velocity of the planet at the equator, which is indicated by the dashed lines ($\pm$5.3 km/s for WASP-76b). When the wind profile is taken into account, the Doppler shifts can be stronger. As shown in Fig. \ref{fig:ccf_maps_full_orbit}, the wind profile causes the spectra to be more blueshifted on average. {\color{black}{In Appendix \ref{app:A}, we plot the phase-dependence of the full width at half maximum (FWHM) for all species and all models in pre- and post-eclipse.}}  

In pre- and post-transit, the behavior of the Doppler shift depends on whether a species also probes the nightside. For Fe and OH, the CCF continues to track the average line-of-sight velocity of the dayside regions that are in view. For CO and H$_2$O, however, the trail ``jumps'' from the dayside to the nightside once the absorption features in the spectrum (which produce negative CCF values) are stronger than the emission features. We will further explore the nightside signals of CO and H$_2$O in Section \ref{subsec:nightside_ccf}.

{\color{white}{---}} \\

\subsection{$K_{\text{p}}$--$V_{\text{sys}}$ maps in pre- and post-eclipse}

Fig. \ref{fig:kp_vsys_maps_dayside} shows the $K_{\text{p}}$--$V_{\text{sys}}$ maps for the pre- and post-eclipse phases of the drag-free model. The maps were obtained by applying equation \ref{eq:kpvsys_sum} to the CCF maps in the left column of Fig. \ref{fig:ccf_maps_full_orbit}. As we consider less than a quarter of the orbit when computing the pre- and post-eclipse maps, we recover the characteristic ``degeneracy'' between $K_{\text{p}}$ and $V_{\text{sys}}$ that is also seen in real datasets (e.g., \citealt{Birkby2013,Nugroho2017,Line2021,Brogi2023}). At both orbital phases, all species show negative $K_{\text{p}}$ offsets between $-$3 and \mbox{$-$7 km/s}. The shifts along the $V_{\text{sys}}$ axis are at most \mbox{1-2 km/s}. When combining the signals from the pre- and post-eclipse (right column in Fig. \ref{fig:kp_vsys_maps_dayside}), we observe two changes. Firstly, the $K_{\text{p}}$--$V_{\text{sys}}$ degeneracy disappears, as we are sampling a larger part of the orbit. Secondly, the only significant peak offsets that remain are those along the $K_{\text{p}}$ axis. The value of $\Delta V_{\text{sys}}$ is negligible in each of the four maps. This is what we expected based on Fig. \ref{fig:ccf_maps_full_orbit}: $\Delta V_{\text{sys}}$ is the \emph{offset} of the trail at the eclipse ($\phi = 180^\circ$), which lies close to zero in each scenario. $\Delta K_{\text{p}}$, however, describes the \emph{rate of change} of the Doppler shifts with orbital phase. Its value is negative because the planet trail goes from being blueshifted to being redshifted in the planet rest frame, counteracting the Doppler shifts due to orbital motion.

\begin{figure*}
\vspace{-10pt}
\makebox[\textwidth][c]{\hspace{30pt}\includegraphics[width=1.15\textwidth]{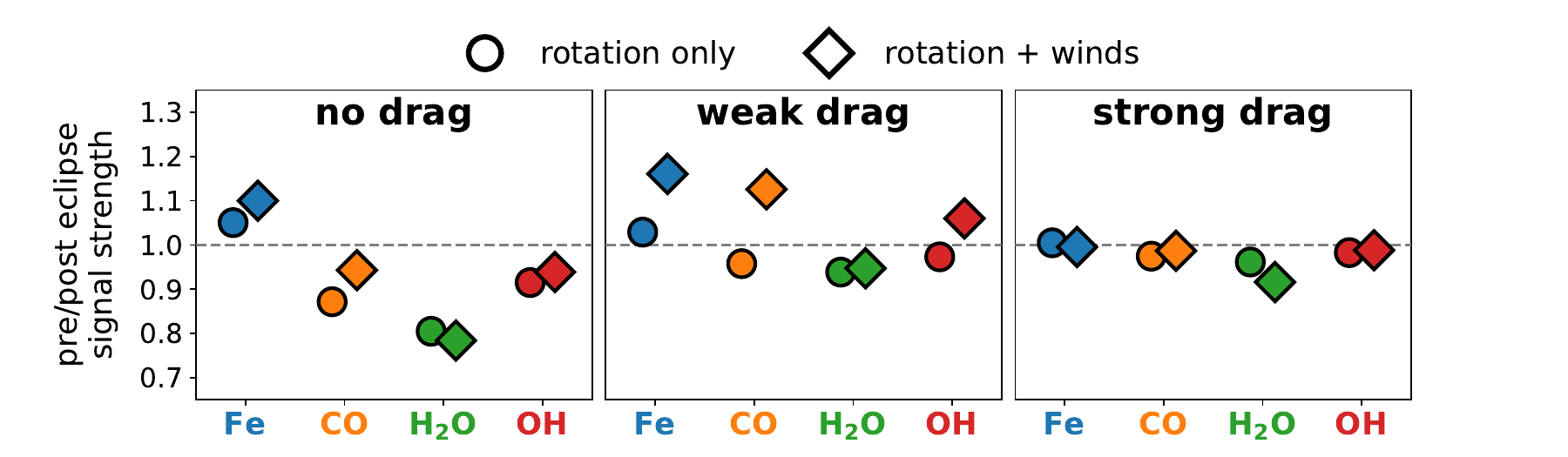}}
\vspace{-18pt}
\caption{Ratios between the maxima of the $K_{\text{p}}$--$V_{\text{sys}}$ maps associated with the pre- and post-eclipse (e.g., Fig. \ref{fig:kp_vsys_maps_dayside}). Each panel corresponds to a different model. The circle markers show the ratio between the signal strengths when only accounting for Doppler effects due to planet rotation. The diamond markers show how the ratio changes when also considering Doppler effects due to the planet's wind profile. Notably, the winds in the weak-drag model cause the CO and OH signals to appear relatively stronger \emph{before} the eclipse, wiping out the signature of the 3D temperature profile.}  
\label{fig:pre_over_post_strength}
\end{figure*}

Fig. \ref{fig:blob_distribution} shows the $K_{\text{p}}$--$V_{\text{sys}}$ offsets that we find for the full set of models. Again, all species show a negative $K_{\text{p}}$ offset, driven by planet rotation. Looking at the combined pre- and post-eclipse signals (indicated by the circle markers), we note that the drag-free model produces stronger $K_{\text{p}}$ offsets than the models that include drag. This is because the drag-free model has an equatorial jet that acts in the same direction as the tidally locked solid-body rotation (see Fig.  \ref{fig:globes_vlos}). Also, within the drag-free model, the $\Delta K_{\text{p}}$ value for Fe \mbox{(ca. $-$5 km/s)} is larger than the $\Delta K_{\text{p}}$ value for H$_2$O (ca. $-$3 km/s). This is because Fe lines probe \emph{lower} pressures on the dayside where H$_2$O is thermally dissociated. At these lower pressures, the wind profile is substantially different (top row in Fig. \ref{fig:globes_vlos}), with the dayside being more blueshifted in pre-eclipse and more redshifted in post-eclipse. In the models with drag and rotation only, the $\Delta K_{\text{p}}$ values of all species are clustered around $-$2-3 km/s, which is roughly half the equatorial rotation velocity of WASP-76b. Finally, we note that $\Delta V_{\text{sys}}$ never exceeds $\pm2$ km/s for any species in any of the explored scenarios. 

In Fig. \ref{fig:pre_over_post_strength} we show the ratio between the peak strengths of the $K_{\text{p}}$--$V_{\text{sys}}$ maps in pre- and post-eclipse for all combinations of GCM outputs and species. The left panel conveys the same information as Fig. \ref{fig:phase_curve}: in the drag-free model, the Fe emission lines are strongest \emph{before} the eclipse, while the line strengths of CO, H$_2$O and OH peak \emph{after} the eclipse. However, as the drag strength increases, the dayside temperature structure becomes more symmetric around the substellar point (e.g., Fig. \ref{fig:equatorial_PT}), and the ratio between the peak strengths should converge to unity for all species. We recover this behavior when only considering Doppler effects due to planet rotation. Yet, if we also take Doppler effects due to the wind profile into account, the picture becomes somewhat more nuanced. In the drag-free and strong-drag models, the wind profile does not significantly affect the ratio between the signal strengths. However, in the weak-drag model, the wind profile causes the CO and OH lines to become relatively stronger \emph{before} the eclipse. This implies that the dispersion of the wind speeds probed during post-eclipse must be larger compared to pre-eclipse, such that the emission lines are subject to stronger Doppler broadening, which lowers the line strength after the eclipse. To summarize, Fig. \ref{fig:pre_over_post_strength} highlights that the 3D temperature structure and the wind profile of a planet can impact the phase-dependent line strengths in opposite ways, warranting caution when it comes to the interpretation of observations.

\subsection{Estimating the $K_{\text{p}}$ offset of the dayside emission signal due to planet rotation}
\label{subsec:kp_estimate}

Our finding that the planet signal should occur at lower $K_{\text{p}}$ values compared to the ``expected'' orbital velocity\footnote{The Keplerian orbital velocity of a planet on a circular orbit is given by $K_{\text{p,orb}} = \frac{2\pi a}{P} \sin(i)$, where $a$ is the semi-major axis, $P$ is the orbital period, and $i$ is the orbital inclination.} $K_{\text{p,orb}}$ is in agreement with high-resolution emission observations of different UHJs such as WASP-18b (\citealt{Brogi2023}), WASP-33b (\citealt{Cont2022}), and WASP-121b (\citealt{Hoeijmakers2022,Bazinet2025}), as well as the JWST/NIRSpec\footnote{The maximum spectral resolution of NIRSpec is about 3,000, which is high enough to detect an atmosphere using cross-correlation (see also \citealt{Esparza2023}).} phase curve of WASP-121b (\citealt{Sing2024}). When it comes to WASP-76b, \citet{costasilva2024} did \emph{not} find a $K_{\text{p}}$ offset for Fe in pre- and post-eclipse spectra from ESPRESSO. Instead, they reported a blueshift along the $V_{\text{sys}}$ axis of $-$4.7$\pm$0.3 km/s (we will further discuss this observation in Section \ref{subsec:vsys_offset}). On the other hand, \citet{Yan2023} did report negative $K_{\text{p}}$ offsets for CO and H$_2$O, albeit with large error bars.

\begin{figure*}
\vspace{-5pt} 
\makebox[\textwidth][c]{\hspace{70pt}\includegraphics[width=1.1\textwidth]{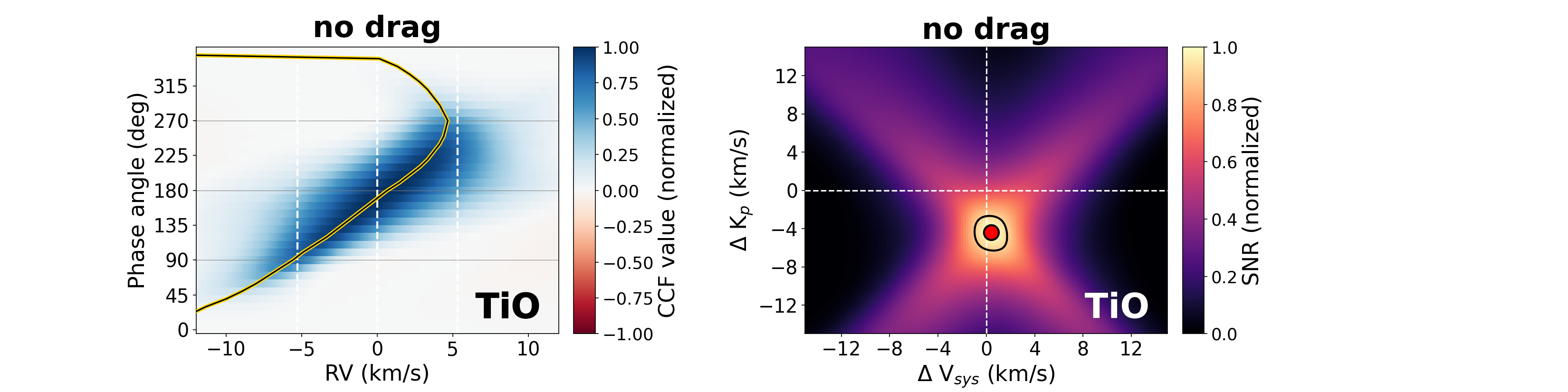}}
\vspace{-17pt}
\caption{CCF map (left panel) and \mbox{$K_{\text{p}}$--$V_{\text{sys}}$} map (right panel) obtained when cross-correlating the drag-free WASP-76b GCMs with a dayside TiO template. As in the right column of Fig. \ref{fig:kp_vsys_maps_dayside}, the \mbox{$K_{\text{p}}$--$V_{\text{sys}}$} map was obtained by combining the signals from the pre- and post-eclipse. Overall, we find that the behavior of the TiO signal in our models is similar to that of Fe and OH. Since there is no significant dissociation of TiO around the hotspot of our WASP-76b models, we find a negative $K_{\text{p}}$ offset.}  
\label{fig:kp_appendix}
\end{figure*}

\citet{Hoeijmakers2022} detected different metals in the atmosphere of WASP-121b at $\Delta K_{\text{p}} \approx -5$ km/s (note that the equatorial rotation velocity of WASP-121b is about 2 km/s higher than that of WASP-76b). To interpret this observation, they presented an analytical model to calculate Doppler shifts associated with points on the equator (assuming an edge-on orbit, \mbox{$V_{\text{sys}}$ =} 0 km/s, and no winds): 

\begin{equation} \label{eq:eq_hoeijmakers}
    v_{\text{los}}(\phi_t,\theta) = \frac{2 \pi}{P} \big[ a \sin(\phi_t) - R_\text{p} \sin(\phi_t-\theta) \big].
\end{equation}

\noindent In this equation, $v_{\text{los}}$ is the line-of-sight velocity at a point with longitude $\theta$ at orbital phase $\phi_t$. It is a combination of the orbital motion of the planet (first term) and its tidally locked rotation (second term). Furthermore, $P$ is the orbital period, $a$ is the semi-major axis, and $R_\text{p}$ is the planet radius. 

In the limit where all the flux emerges from the substellar point ($\theta = 0^\circ$), the line-of-sight velocity becomes

\begin{equation}
\begin{split}
    v_{\text{los}}(\phi_t) &= \frac{2 \pi (a - R_p)}{P} \sin(\phi_t) 
    = (K_{\text{p,orb}} - v_{\text{eq}})\sin(\phi_t) \\
    &= (K_{\text{p,orb}} + \Delta K_{\text{p}})\sin(\phi_t)
    = K_{\text{p,obs}} \sin(\phi_t),
\end{split}
\end{equation}

\noindent where $K_{\text{p,orb}} = 2\pi a/P$ is the Keplerian orbital velocity of the planet, $v_\text{eq} = 2\pi R_\text{p}/P$ is the rotational velocity at the equator, and $K_{\text{p,obs}}$ is the ``observed'' $K_{\text{p}}$ based on the emission spectrum. In the above scenario, the $K_{\text{p}}$ offset is equal to the equatorial rotation velocity of the planet. In reality, the signal will not only come from the substellar point. Therefore, for a species that is distributed \emph{more or less uniformly} across the dayside, $v_\text{eq}$ should be seen as an upper limit: \mbox{$|\Delta K_{\text{p}}| < v_\text{eq}$} (see also Fig. \ref{fig:blob_distribution}). When also considering the planet's wind profile, the inequality can be generalized to \mbox{$|\Delta K_{\text{p}}| < (v_\text{eq} + v_\text{jet})$}, with $v_\text{jet}$ the equatorial jet speed. We note that these upper limits on  $\Delta K_{\text{p}}$ are much lower than in transmission, where the offsets along the $K_{\text{p}}$ axis can be multiple times the equatorial rotation velocity of the planet\footnote{Equation 8 from \citet{Wardenier2023} yields a $\Delta K_{\text{p}}$ estimate of $\pm$21 km/s for the transmission spectrum of WASP-76b, which is much larger than the equatorial rotation velocity of $\pm$5.3 km/s.} (\citealt{Wardenier2023}).

As demonstrated by \citet{Cont2021}, who detected TiO in the emission spectrum of WASP-33b ($T_{\text{eq}}$ $\sim$ 2700 K) at $\Delta K_{\text{p}} \approx +17$ km/s, there exist certain scenarios in which the $K_{\text{p}}$ offset of a species can be \emph{positive}. When a species is depleted near the hotspot of the planet (e.g., due to thermal dissociation), the strongest signals will emerge from regions near the morning and evening terminators (at $\theta \approx \pm90^\circ$). Consequently, the Doppler shifts pick up an additional redshift/blueshift during pre-/post-eclipse due to planet rotation, leading to a larger apparent orbital velocity (see Figs. 8 and 9 in \citealt{Cont2021}). In our models of WASP-76b, however, we do not encounter this behavior (also not for TiO; see Fig. \ref{fig:kp_appendix}). This is likely because the equilibrium temperature of WASP-76b is 500 K lower than that of WASP-33b.

\begin{figure*}
\vspace{-35pt} 
\makebox[\textwidth][c]{\hspace{-1pt}\includegraphics[width=1.0\textwidth]{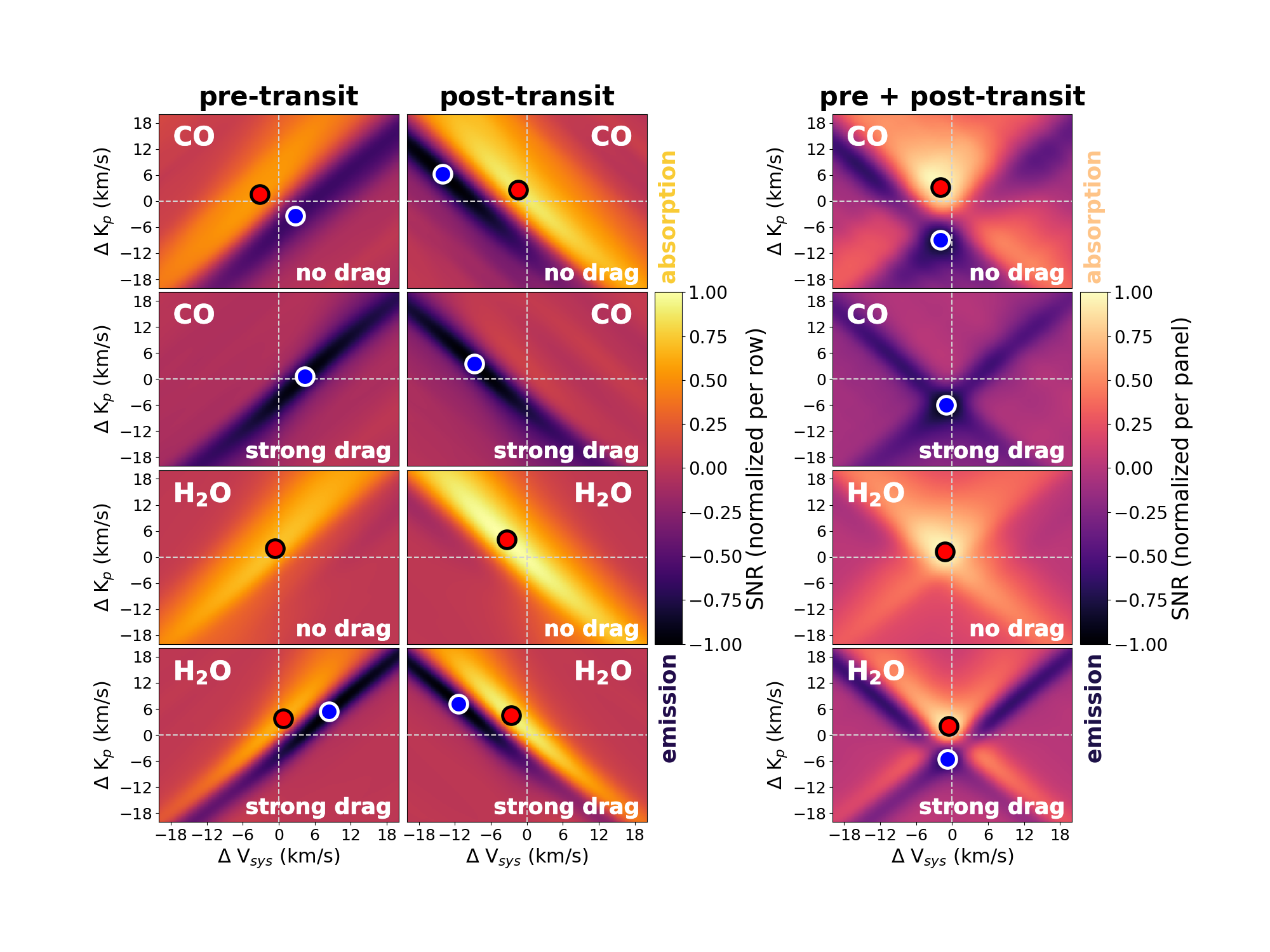}}
\vspace{-45pt}
\caption{Plot similar to Fig. \ref{fig:kp_vsys_maps_dayside}, but now for the pre- and post-transit phases. The figure shows the $K_{\text{p}}$--$V_{\text{sys}}$ maps for CO (top two rows) and H$_2$O (bottom two rows) in the drag-free and the strong-drag models, obtained by cross-correlating the planet spectra with a nightside template. The red and blue markers indicate the locations of the positive and/or negative peaks in each of the maps, respectively. In the absence of Doppler effects, the planet signal would be situated at \mbox{($\Delta$$K_{\text{p}}$, $\Delta$$V_{\text{sys}}$) =} \mbox{(0, 0)} km/s, indicated by the white dashed lines.}
\label{fig:kp_vsys_maps_nightside}
\vspace{0pt}
\end{figure*}

\subsection{$K_{\text{p}}$--$V_{\text{sys}}$ maps in pre- and post-transit}
\label{subsec:nightside_ccf}

As demonstrated in Sections \ref{subsec:spectral_lines} and \ref{subsec:line_strengths}, interpreting the planet signals during pre- and post-transit is more challenging for two main reasons: (i) the signals are much weaker than in pre-/post-eclipse due to the shallower nightside temperature gradient and lower flux levels, and (ii) the spectra contain a mix of absorption and emission features. In theory, one could cross-correlate the planet spectra with phase-dependent templates derived from a 3D model (e.g., \citealt{Beltz2020,Beltz2022}). In reality, however, the true underlying temperature structure of the target is unknown, so the 3D template will never be a perfect representation of the spectrum. Therefore, to keep things simple and intuitive, we will focus on measuring Doppler shifts using a nightside template (see Section \ref{subsec:CCF_description}) that does not depend on orbital phase and only contains absorption lines.

\begin{figure*}
\vspace{-35pt} 
\makebox[\textwidth][c]{\hspace{-1pt}\includegraphics[width=1.17\textwidth]{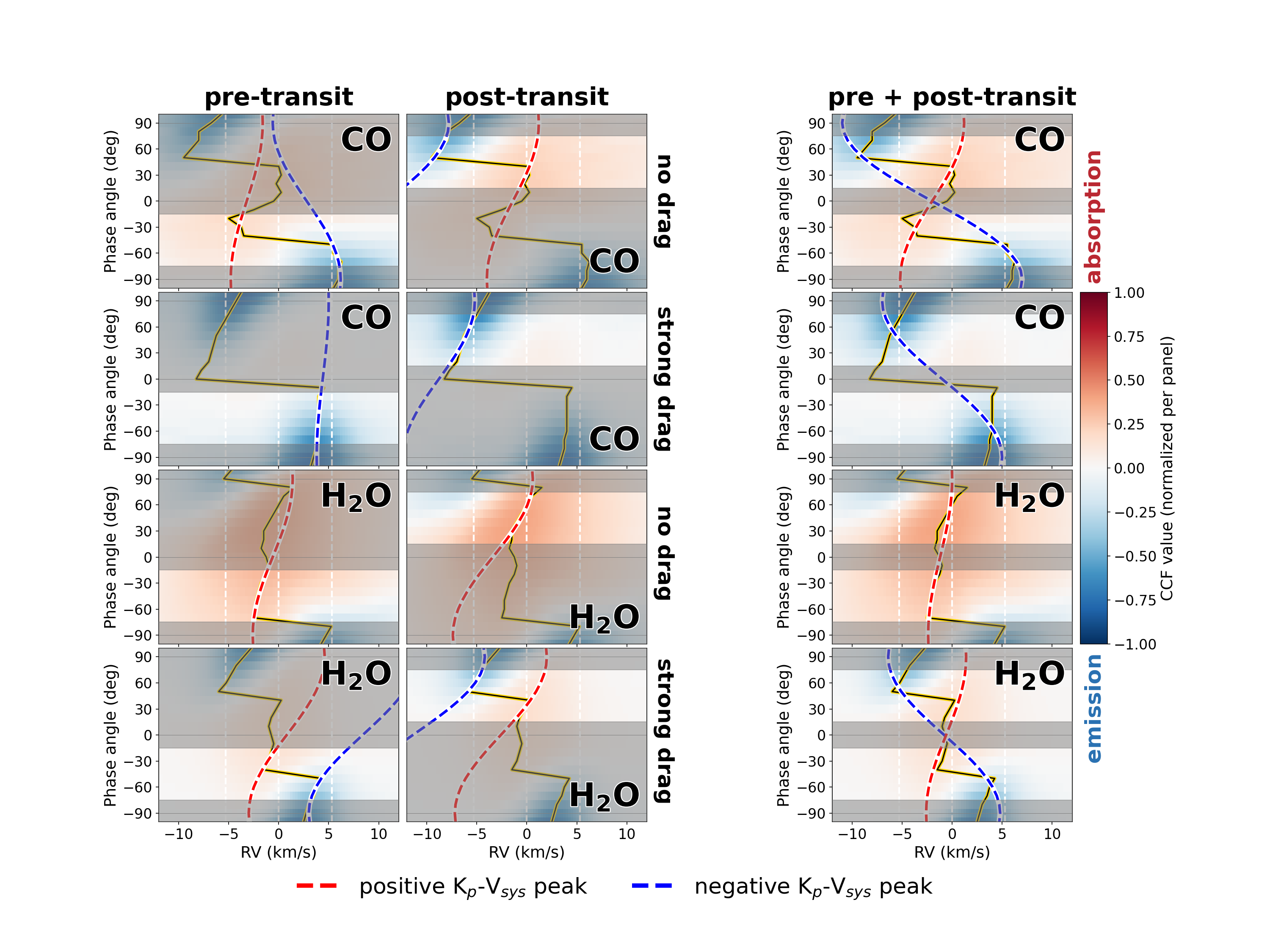}}
\vspace{-35pt}
\caption{Visualization of the ``best-fit'' trails corresponding to the positive and negative peaks in the $K_{\text{p}}$--$V_{\text{sys}}$ maps in Fig. \ref{fig:kp_vsys_maps_nightside}. The rest-frame CCF maps are shown in the background. The CCF maps were obtained by cross-correlating \emph{all} planet spectra with the static nightside templates from the drag-free model. Hence, a positive CCF implies \emph{absorption} features, while a negative CCF implies \emph{emission} features. Each panel highlights the phase ranges over which the $K_{\text{p}}$--$V_{\text{sys}}$ maps were computed, while the rest of the CCF map is shaded in grey. The yellow trails mark the most extreme CCF values at each orbital phase.}
\label{fig:kp_vsys_trails_nightside}
\end{figure*}

Fig. \ref{fig:kp_vsys_maps_nightside} shows the $K_{\text{p}}$--$V_{\text{sys}}$ maps of CO and H$_2$O for the pre- and post-transit phases of WASP-76b when cross-correlating the spectra with a nightside template. In Fig. \ref{fig:kp_vsys_trails_nightside}, we show the CCF maps with the orbital trails corresponding to the $K_{\text{p}}$--$V_{\text{sys}}$ peaks from Fig. \ref{fig:kp_vsys_maps_nightside} plotted on top. The layout of both figures is the same, such that panels in the same rows and columns correspond to each other. Furthermore, we do not discuss the signals of Fe and OH here as these species only probe the dayside of the planet. If detectable during the pre- and post-transit, their emission lines will simply track the dayside regions that are still in view (see Fig. \ref{fig:ccf_maps_full_orbit}).

In the drag-free model, the $K_{\text{p}}$--$V_{\text{sys}}$ maps of CO display both a positive and a negative peak. This is because the underlying CCF map contains both positive and negative values in pre- and post-transit (see first row in Fig. \ref{fig:kp_vsys_trails_nightside}). Therefore, when co-adding the CCF values along the phase axis, there exist two well-defined trails along which the sum of the CCFs is either maximized (capturing nightside absorption features) or minimized (capturing dayside emission features). As illustrated in Fig. \ref{fig:kp_vsys_trails_nightside}, these trails do not adequately describe the true planet signal (in yellow) over the \emph{whole} phase range that is considered -- they only match the signal at very specific phases. In fact, co-adding the CCF maps along the actual planet trail would lead to a weaker aggregate signal, as the positive and negative CCF values will (partly) cancel each other out. Additionally, we note that the structure of our post-transit $K_{\text{p}}$--$V_{\text{sys}}$ map for CO is reminiscent of Fig. 2 from \citet{Mraz2024}, who claim a detection of CO on the nightside of WASP-33b\footnote{If robust, these observations would point to little drag and relatively efficient heat redistribution throughout the atmosphere of WASP-33b.}.

When combining the pre- and post-transit signals of CO, the positive and negative peaks in the $K_{\text{p}}$--$V_{\text{sys}}$ map acquire opposite $K_{\text{p}}$ offsets. This is because the dayside rotates \emph{away} from the observer in pre-transit and \emph{towards} the observer in post-transit ($\Delta K_{\text{p}} < 0$), while the opposite holds for the nightside ($\Delta K_{\text{p}} > 0$). As shown in the right column of Fig. \ref{fig:kp_vsys_trails_nightside}, we need an orbital trail with a positive amplitude (in red dashes) to \emph{maximize} the sum of CCF values, while we need a trail with a negative amplitude (in blue dashes) to \emph{minimize} this sum. However, since the trails are only set by two parameters, neither of them is able to fully reproduce the more intricate phase dependence of the true planet trail.

In the strong-drag model (second row in Figs. \ref{fig:kp_vsys_maps_nightside} and \ref{fig:kp_vsys_trails_nightside}) the $K_{\text{p}}$--$V_{\text{sys}}$ map of CO only features a negative peak, indicating that dayside emission features are dominant across the whole pre- and post-transit phase. This behavior is in agreement with the phase dependence of the line strengths plotted in Fig. \ref{fig:line_contrast_curves}. In this scenario, we are thus only probing the dayside of WASP-76b.

In contrast to the CO signal, the H$_2$O lines in the drag-free model are visible in absorption during the entire pre- and post-transit (third row in Figs. \ref{fig:kp_vsys_maps_nightside} and \ref{fig:kp_vsys_trails_nightside}). As a result, the $K_{\text{p}}$--$V_{\text{sys}}$ map only contains a positive peak. Furthermore, because the (yellow) planet trail does not suffer from discontinuities or sign changes of the CCF values, the $K_{\text{p}}$ and $V_{\text{sys}}$ values associated with the peak are a very good description of the actual planet signal.

In the strong-drag model (bottom row in Figs. \ref{fig:kp_vsys_maps_nightside} and \ref{fig:kp_vsys_trails_nightside}), the behavior of the H$_2$O signal is very similar to that of the CO signal in the drag-free model. Because the nightside of this model is more isothermal, H$_2$O absorption is limited to a narrower range of phases. Hence, the $K_{\text{p}}$--$V_{\text{sys}}$ maps feature two peaks again, which are associated with the dayside and the nightside, respectively.

\begin{figure*}
\vspace{-20pt} 
\makebox[\textwidth][c]{\hspace{10pt}\includegraphics[width=0.7\textwidth]{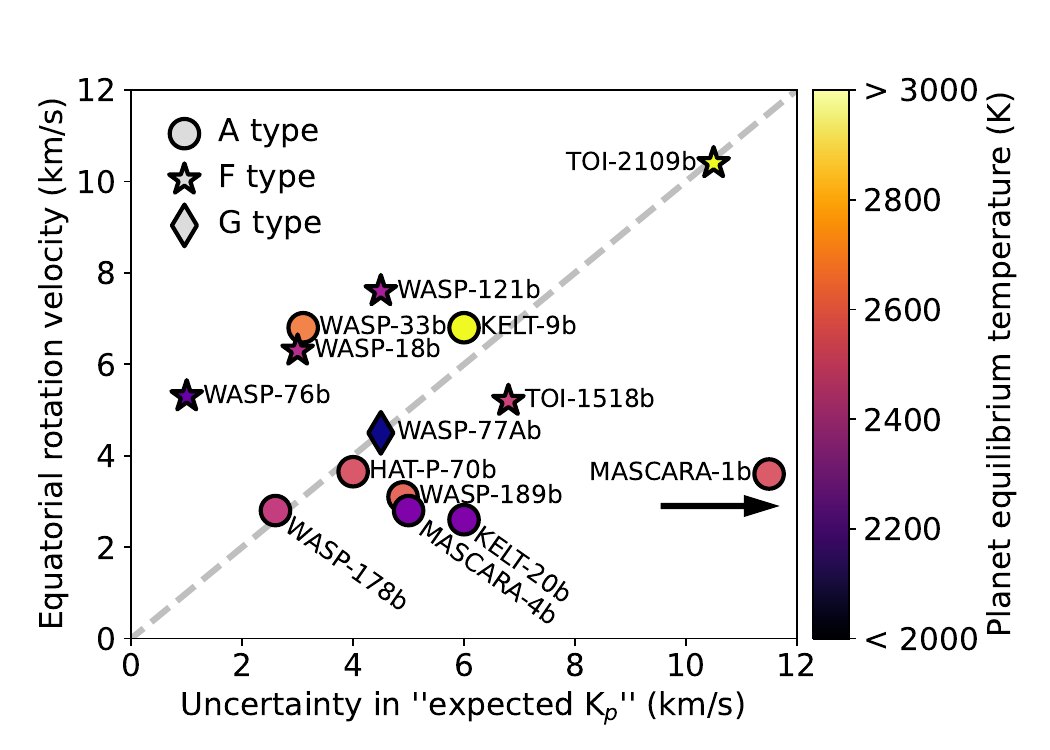}}
\vspace{-22pt}
\caption{The equatorial rotation velocity of different UHJs plotted against the uncertainty in the literature value of their Keplerian orbital velocity $K_{\text{p,orb}}$. This value is often referred to as the ``expected $K_{\text{p}}$'' of the planet. The marker shapes denote the classification of the host star. Furthermore, the color scale indicates the planetary equilibrium temperature. As demonstrated in Section \ref{subsec:kp_estimate}, the equatorial rotation velocity is a rough upper bound for the $K_{\text{p}}$ offsets seen in pre- and post-eclipse spectra. For all planets \emph{below} the dashed line, the current uncertainty in the Keplerian orbital velocity is larger than the $K_{\text{p}}$ offsets that planet rotation will imprint on their emission spectra. References for the $K_{\text{p,orb}}$ uncertainties are as follows: WASP-76b, WASP-121b, WASP-189b, KELT-20b, MASCARA-4b, HAT-P-70b (\citealt{Gandhi2023}); WASP-33b (\citealt{Cont2022}); WASP-18b (\citealt{Yan2023}); TOI-1518b (\citealt{Simonnin2024}); WASP-77Ab (\citealt{Smith2023}); WASP-178b (\citealt{Damasceno2024}); \mbox{MASCARA-1b} (25 km/s; \citealt{Ramukar2023}). The uncertainties for KELT-9b and TOI-2109b were derived from the semi-major axis and orbital period values/errors reported in \citet{Stangret2024} and \citet{Wong2021}, respectively.} 
\label{fig:kp_uncertainty}
\end{figure*}

A final aspect to highlight is that the negative $K_{\text{p}}$--$V_{\text{sys}}$ peaks (associated with dayside emission) in the individual pre- and post-transit maps tend to have strong $V_{\text{sys}}$ offsets. This is because, mathematically, $\Delta V_{\text{sys}}$ is the radial velocity at which the corresponding trail intersects the line $\phi = 0^\circ$ in the CCF map. The negative peaks, however, encode information from the dayside, whose signal gets stronger at phases further away from $\phi = 0^\circ$. Therefore, the $\Delta V_{\text{sys}}$ value should not be interpreted as an effective line-of-sight wind speed. For example, in the post-transit map of H$_2$O in the strong-drag model, the Doppler shift of the dayside is about \mbox{$-$5 km/s}. However, the trail that corresponds to the negative peak in the \mbox{$K_{\text{p}}$--$V_{\text{sys}}$} map intersects the line $\phi = 0^\circ$ at $\Delta V_{\text{sys}} \approx -12$ km/h.

\subsection{A note on measuring $K_{\text{p}}$ offsets from high-resolution observations}

To infer the $K_{\text{p}}$ offset of a species from a \mbox{$K_{\text{p}}$--$V_{\text{sys}}$} map, we need a reference value for the true (``expected'') orbital velocity $K_{\text{p,orb}}$ of the planet. So far, we have assumed that our knowledge of this value is perfect. However, as shown in Fig. \ref{fig:kp_uncertainty}, the uncertainties in the literature values of $K_{\text{p,orb}}$ are typically multiple km/s for UHJs. As discussed in Section \ref{subsec:kp_estimate}, we expect the $K_{\text{p}}$ offset of a species in the dayside spectra to be of similar magnitude. Therefore, it will be hard for certain planets to meaningfully measure $K_{\text{p}}$ offsets in emission, unless the errors in their true orbital velocities are reduced.

In general, the Keplerian orbital velocity $K_{\text{p,orb}}$ of a planet should obey the following equation (\citealt{Torres2010,Sing2024}):

\begin{equation}
\begin{split}
    M_{*} \sin^3(i) = \xi (1 - e^2)^{3/2} (K_{*}+K_\text{p,orb})^2 K_\text{p,orb} P,
\end{split}
\end{equation}

\noindent where $M_{*}$ is the stellar mass, $i$ is the orbital inclination, $\xi = 1.036149 \times 10^{-7}$ is a constant, $e$ is the orbital eccentricity, $K_{*}$ is the stellar RV semi-amplitude, and $P$ is the orbital period. To lower the uncertainty in $K_\text{p,orb}$, we thus need to reduce the error in the other parameters in this equation.    

For UHJs, the eccentricity, inclination angle, and orbital period tend to be very well constrained (e.g., \citealt{Ehrenreich2020,Borsa2021,Wong2021}). Based on dynamical arguments (\citealt{Rasio1996}), it is often assumed that $e = 0$. So far, observations have been consistent with this assumption within tight error margins. Furthermore, the errors in the orbital period of UHJs are typically less than a second. For transiting planets, the orbital inclination can be constrained from the Rossiter-McLaughlin effect. 

This leaves us with $M_{*}$ and $K_{*}$. For a single star, the way to infer its mass is through evolutionary tracks and isochrones (e.g., \citealt{Schaller1992,Torres2010,Sing2024}). Hence, the error in $M_{*}$ is impacted by model uncertainties, as well as measurement uncertainties of the star's effective temperature, surface gravity, and metallicity. For WASP-76 and WASP-121, for example, the error bar on the stellar mass is about \mbox{0.02 $M_\odot$} (\citealt{Tabernero2020,Borsa2021}). 

The value of $K_{*}$ can be constrained from radial-velocity measurements. The main reason why the $K_{\text{p,orb}}$ uncertainty for WASP-76b is so low is the small error bar on the $K_{*}$ value of its host star ($116.02^{+1.29}_{-1.35}$ m/s; \citealt{Ehrenreich2020}). For stars such as WASP-121 ($K_{*} = 177 \pm 8$ m/s; \citealt{Bourrier2020}) and TOI-2109 ($K_{*} = 860 \pm 130$ m/s; \citealt{Wong2021}), the error bars are significantly larger, resulting in a higher uncertainty in the expected planetary orbital velocity. 

The relatively large uncertainties in the $K_{\text{p,orb}}$ values of some UHJs could be a motivation for more in-depth characterization of their host stars to better constrain their RV semi-amplitude. High-precision spectrographs such as ESPRESSO, MAROON-X, or NIRPS would be ideal for this task. We note, however, that A-type stars such as WASP-33 and WASP-189 are fast-rotating and/or active, which could make it significantly harder to lower the $K_{*}$ error to the km/s level.

Alternatively, a way to circumvent the observational uncertainties in $K_{\text{p,orb}}$ could be to report differences between the $\Delta K_\text{p}$ values of two species (say, Fe and H$_2$O), thereby ``subtracting out'' the planet's orbital velocity. {\color{black}{However, as shown Fig. \ref{fig:blob_distribution}, differences between the $\Delta K_\text{p}$ values of two species will typically be small (at most 1-2 km/s in the drag-free model), so these measurements would require high precision to be meaningful.}}  

\subsection{A note on $V_{\text{sys}}$ offsets}
\label{subsec:vsys_offset}

Our WASP-76b models do not predict any large $V_{\text{sys}}$ offsets in the planet's dayside emission spectrum, especially when pre- and post-eclipse observations are combined. The reason for this is that the CCF trails in Fig. \ref{fig:ccf_maps_full_orbit} intersect the line $\phi = 180^\circ$ at RV $\approx$ 0 km/s, which is best fit by an orbital trail with $\Delta V_{\text{sys}} \approx 0$ km/s. In light of our models, the consistent blueshifts observed in the Fe lines of WASP-76b during pre- \emph{and} post-eclipse with ESPRESSO \mbox{($\Delta V_{\text{sys}} =$} $-4.7 \pm 0.3$ km/s; \citealt{costasilva2024}) are a bit puzzling\footnote{{\color{black}{Just before submitting the revised version of this manuscript, we learned that a re-analysis of the ESPRESSO data by another team revealed no shifts in $V_\text{sys}$, but rather a negative $\sim$5 km/s shift in $K_{\text{p}}$. This result is much more in agreement with our models (G. Guilly, private communication).}}}. \citet{costasilva2024} suggested that this blueshift could be caused by a strong updraft of material around the substellar point of the planet (see their Fig. 8), which effectively moves Fe towards the observer. Our GCMs, however, do not provide strong evidence for such vertical transport, perhaps due to missing physics\footnote{One mechanism that our WASP-76b models do not consider is heat transport due to hydrogen dissociation/recombination (e.g., \citealt{Bell2018,Komacek2018,Tan2019,Roth2021}). However, post-processing of the WASP-121b models from \citet{Wardenier2024} (which include hydrogen dissociation/recombination) shows that this additional physics does not lead to stronger $V_{\text{sys}}$ offsets in dayside emission spectra. We refer to Appendix B in \citet{Bazinet2025} for the \mbox{$K_{\text{p}}$--$V_{\text{sys}}$} maps of these models.}. Additional forward modeling and follow-up observations with different instruments will be required to resolve the discrepancies between the current data and our models. 

One aspect of the observations from \citet{costasilva2024} that is more in agreement with our models is the \emph{decreasing} blueshift of Fe over the course of the orbit: RV = $-$6.0 $\pm$ 0.4 km/s in pre-eclipse, while RV = $-$3.3 $\pm$ 0.5 km/s in post-eclipse. This could be a signature of planet rotation. In principle, such a change in Doppler shift with phase should translate to a (negative) $K_{\text{p}}$ offset in the \mbox{$K_{\text{p}}$--$V_{\text{sys}}$} map. However, \citet{costasilva2024} reported that the Fe emission signal of WASP-76b is consistent with $\Delta K_{\text{p}} = 0$ km/s. One reason for this could be differences in the signal-to-noise ratio between the four ESPRESSO visits, such that not all orbital phases contribute equally to the ``\mbox{$K_{\text{p}}$--$V_{\text{sys}}$} fit''. 

While \citet{costasilva2024} found a \emph{negative} $V_{\text{sys}}$ offset for Fe on WASP-76b, \citet{Lesjak2024} reported \emph{positive} $V_{\text{sys}}$ offsets for Fe and CO on WASP-189b, based on pre- and post-eclipse observations from CRIRES+ ($\Delta V_{\text{sys}} \approx +6$ km/s). \citet{Lesjak2024} attributed their measurements to strong day-to-night winds that impart a redshift on the emission spectrum. With regard to WASP-189b, we note that its equatorial rotation velocity is about 2 km/s lower than that of WASP-76b (see Fig. \ref{fig:kp_uncertainty}), which leads to smaller $K_{\text{p}}$ offsets due to planet rotation. In this case, it is likely that the wind profile is the dominant driver of the line-of-sight velocities. Along this line of reasoning, one could speculate that UHJs with high rotation rates (i.e. short orbital periods) are more likely to show $K_{\text{p}}$ offsets, while UHJs with lower rotation rates are likely to show stronger $V_{\text{sys}}$ offsets. Such trends could be identified by modeling emission spectra for a sample of UHJs.     

\section{Summary and conclusion}
\label{sec:conclusion}

In this work, we simulated phase-dependent emission spectra of the canonical ultra-hot Jupiter (UHJ) \mbox{WASP-76b}, based on 3D atmospheric models from the SPARC/MITgcm. We post-processed the GCMs with gCMCRT to obtain emission spectra at 36 phases along the orbit. We then cross-correlated the spectra with different templates to obtain CCF and \mbox{$K_{\text{p}}$--$V_{\text{sys}}$} maps for Fe, CO, H$_2$O, and OH. This allowed us to identify how their Doppler shifts and line strengths evolve throughout the orbit. Because of the stark day-night temperature contrasts that prevail on UHJs, the emission spectra undergo drastic changes as different parts of the 3D temperature structure rotate into and out of view. 

As the number of high-resolution observations of UHJs is steadily increasing, 3D forward-modeling studies are crucial to fully leverage the information content of current and future telescope data. In a few years' time, UHJs will undoubtedly be prime targets for the E-ELT (e.g., \citealt{Palle2023}), which will deliver transmission and emission spectra at unprecedented signal-to-noise ratios. These will allow us to characterize the climate and composition of UHJs in much finer detail. Finally, we note that cross-correlation analyses of UHJs are now also possible with JWST/NIRSpec (\citealt{Sing2024}), albeit at more modest spectral resolutions ($R \sim 3,000$). This means that our work will also be relevant for understanding certain aspects of space-based observations.

We summarize our most important findings below:

\begin{enumerate}

\item[$\bullet$] Detecting the nightsides of UHJs (through CO and H$_2$O absorption lines) could be an order of magnitude harder than detecting their daysides. This is due to the more isothermal temperature profile of the nigthside, which causes the absorption features to be much weaker than the emission features associated with the (thermally inverted) dayside. In our models with drag, parts of the pre- and post-transit phases are dominated by dayside emission, even though a much larger portion of the nightside is in view. In general, the detectability of the nightside spectrum will depend on the efficiency of heat redistribution across the atmosphere of an UHJ, which sets the vertical temperature gradient on the nightside.  

\item[$\bullet$] Line strengths in the thermal emission spectrum of a planet are set by the vertical temperature gradient and not by the total flux. For example, our drag-free model of WASP-76b has an eastward hotspot offset (such that the total flux peaks before the eclipse), while the line strengths of CO, H$_2$O, and OH are highest after the eclipse. 

\item[$\bullet$] In each of our WASP-76b models, the phase-dependent Doppler shifts show a ``quasi-sinusoidal'' behavior in the planet rest frame that is driven by planet rotation. During pre- and post eclipse, this results in a negative $K_{\text{p}}$ offset in the \mbox{$K_{\text{p}}$--$V_{\text{sys}}$} map of a species. The upper limit on this offset is given by \mbox{$|\Delta K_{\text{p}}| < (v_\text{eq} + v_\text{jet})$}, with $v_\text{eq}$ the equatorial rotation velocity of the planet and $v_\text{jet}$ the equatorial jet speed (in the case of atmospheric drag the last term can be ignored). When combining the pre- and post-eclipse phases, we find negligible $V_{\text{sys}}$ offsets (at most 1 km/s) for our WASP-76b models.   

\item[$\bullet$] In pre- and post-transit, CO and H$_2$O can appear in emission \emph{and} absorption. If this is the case, the corresponding \mbox{$K_{\text{p}}$--$V_{\text{sys}}$} maps will feature both a positive and a negative peak. When combining the pre- and post-transit phases, the peaks will be located at opposite \mbox{$K_{\text{p}}$} offsets that are $\lesssim 2v_\text{eq}$ apart. In the individual \mbox{$K_{\text{p}}$--$V_{\text{sys}}$} maps of the pre- and post-transit, the peaks can have have larger $V_{\text{sys}}$ offsets that are not necessarily a measure for the line-of-sight velocities in the atmosphere.

\item[$\bullet$] For many UHJs the uncertainty in the true Keplerian orbital velocity $K_\text{p,orb}$ is roughly as big as the $K_\text{p}$ offset which their atmospheres could imprint on the emission spectra. If this is the case, it will be hard to meaningfully constrain $\Delta K_{\text{p}}$ values and make inferences about the dynamics of the planet. More in-depth RV characterization of UHJ host stars would be beneficial in this regard.

\end{enumerate}

\vspace{10pt}

\noindent J.P.W. acknowledges support from the Trottier Family Foundation via the Trottier Postdoctoral Fellowship, held at the Institute for Research on Exoplanets (IREx). We also  acknowledge support from the Canadian Space Agency (CSA) under grant 24JWGO3A-03. V.P. is funded by the French National Research Agency (ANR) project EXOWINDS (ANR-23-CE31-0001-01). We thank the members of the NIRPS science team for insightful discussions.


%

\vspace{5mm}
\facilities{--}


\software{\texttt{matplotlib} (\citealt{Hunter2007}), \texttt{numpy} (\citealt{Harris2020}), \texttt{scipy} (\citealt{Virtanen2020}), gCMCRT (\citealt{Lee2022}), SPARC/MITgcm (\citealt{Showman2009})}





\appendix

{{\color{black}{

\section{Supplementary figures}
\label{app:A}

In this appendix, we present two supplementary figures. \textbf{Figure \ref{fig:one_dimensional_contribution}} shows the rough pressures probed by the emission lines of Fe, CO, H$_2$O, and OH, as derived from the 1D hemispheric dayside average of the drag-free GCM of \mbox{WASP-76b} (see the figure caption for further information about methods). The plots demonstrate that different species probe the atmosphere at a range of pressures between $\sim$10$^{-1}$ and $\sim$10$^{-5}$ bar. While H$_2$O probes relatively deep layers of the atmosphere (pressures higher than $\sim$10$^{-3}$ bar), the Fe emission lines mostly emerge from pressures between $\sim$10$^{-3}$ and $\sim$10$^{-5}$ bar. 

\textbf{Figure \ref{fig:fwhm_plot}} shows the full width at half maximum (FWHM) of the CCFs from Fig. \ref{fig:ccf_maps_full_orbit}. During pre-eclipse ($\phi < 180^\circ$), the FWHM increases for all species as a larger part of the dayside rotates into view. That is, the emission signal emerges from an increasingly larger region on the planet disk, probing a wider range of line-of-sight velocities. The opposite effect occurs during post-eclipse ($\phi < 180^\circ$), when parts of the dayside rotate out of view. Comparing the different GCM outputs of WASP-76b, we note that the drag-free model produces the broadest CCF peaks. This is because its strong winds, including the equatorial jet, increase the dispersion in the probed line-of-sight velocities \mbox{(see Fig. \ref{fig:globes_vlos})}. Interestingly, at $\phi = 180^\circ$, the FWHMs of the weak-drag model are $\sim$2 km/s lower than the FWHMs of the strong-drag model, suggesting that the day-to-night flow partially counteracts the line-of-sight velocities due to planet rotation. For CO and Fe, the weak-drag model also predicts the CCF peaks to be somewhat broader in post-eclipse than in pre-eclipse.

\begin{figure*}
\vspace{-20pt} 
\makebox[\textwidth][c]{\hspace{20pt}\includegraphics[width=0.8\textwidth]{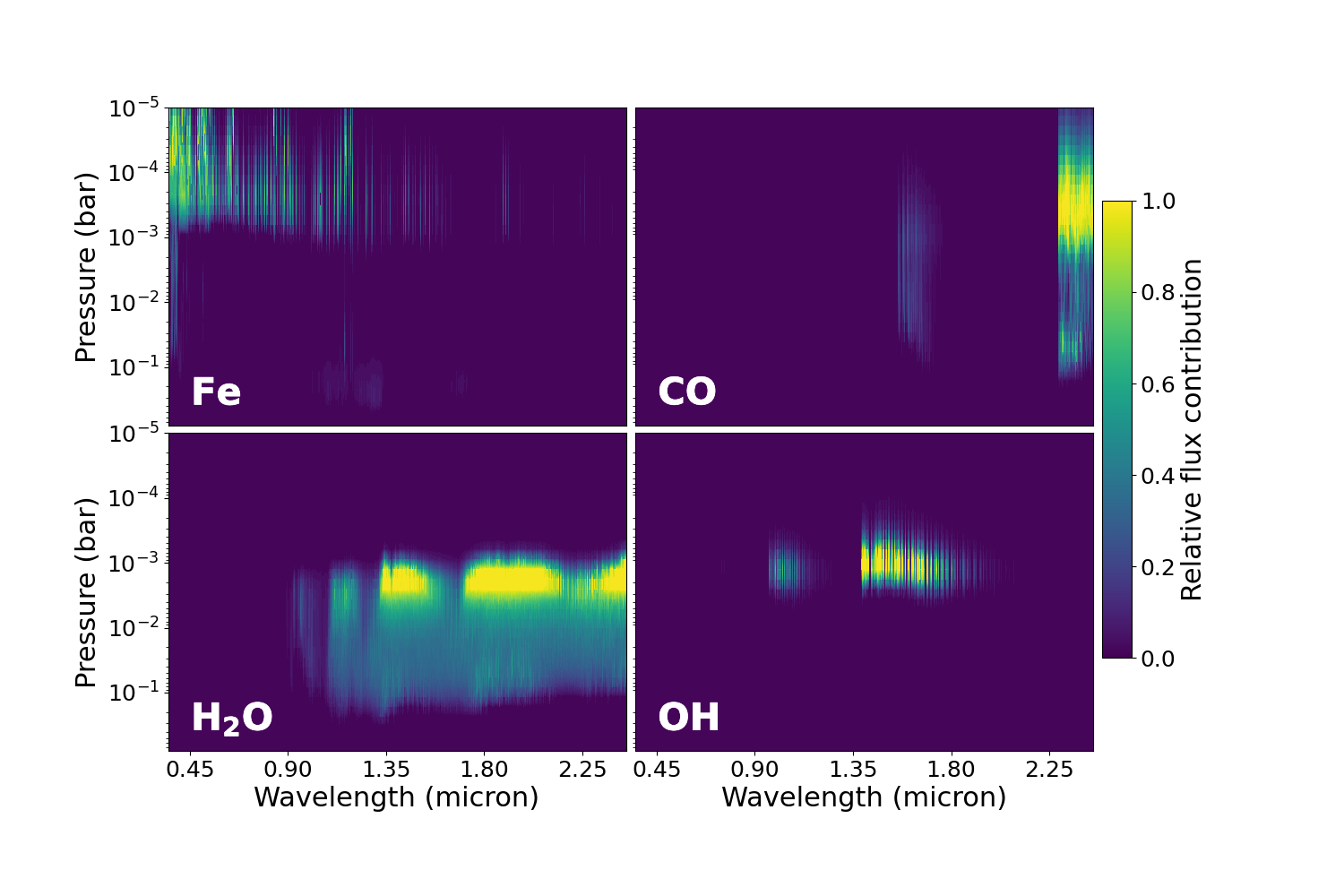}}
\vspace{-35pt}
\caption{\color{black}{1D emission contribution plots generated with petitRADTRANS (\citealt{Molliere2019a}) highlighting the rough pressures probed by Fe, CO, H$_2$O, and OH. The underlying 1D atmospheric model was obtained by averaging the dayside temperature profile of the drag-free GCM output for WASP-76b and recomputing chemical abundances using easyCHEM (\citealt{Lei2024}). To obtain the contribution plot for an individual species $X$, we took the difference between the contribution function from the model with all species and the contribution function from the model with \emph{all species except $X$}. Because the contribution functions output by petitRADTRANS are normalized by construction, taking their difference also resulted in (small) negative values at certain pressures and wavelengths. For clarity, these were set to zero.}}
\label{fig:one_dimensional_contribution}
\end{figure*}

\begin{figure*}
\vspace{-10pt} 
\makebox[\textwidth][c]{\hspace{20pt}\includegraphics[width=1.2\textwidth]{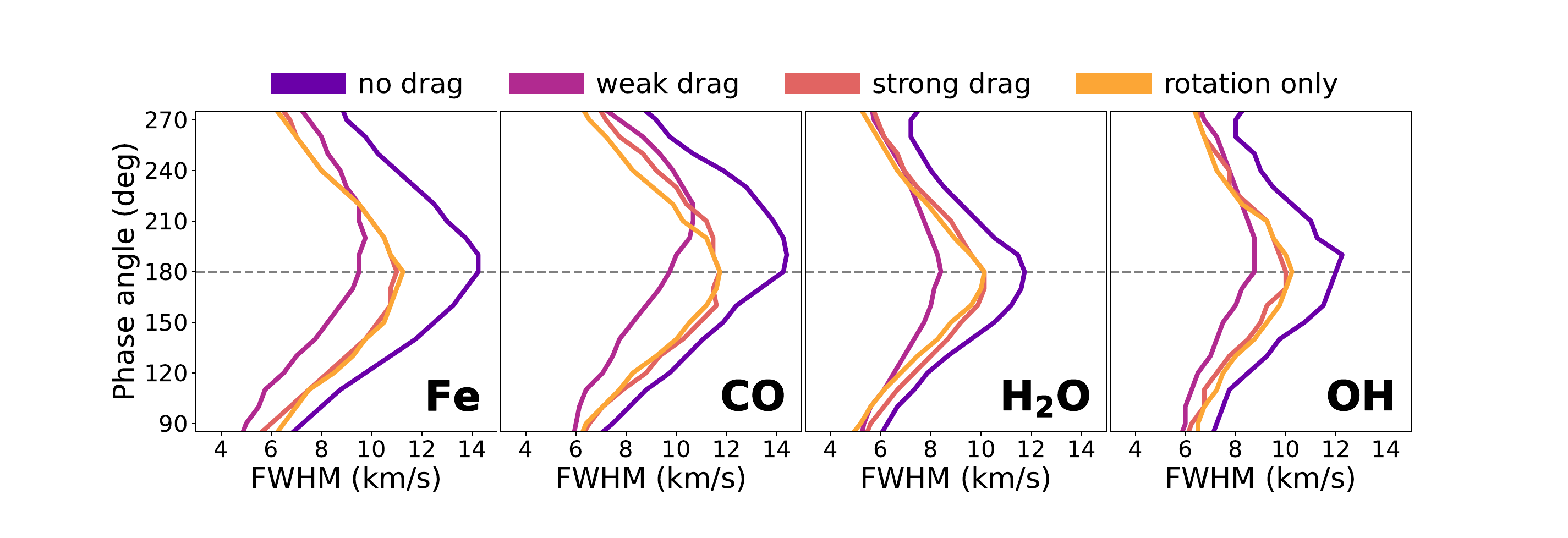}}
\vspace{-25pt}
\caption{\color{black}{Phase dependence of the full width at half maximum (FWHM) of the CCFs plotted in Fig. \ref{fig:ccf_maps_full_orbit}. Each panel corresponds to a different species, while each curve corresponds to a different GCM output (see legend). The grey dashed lines indicate the secondary eclipse.}}
\vspace{15pt}
\label{fig:fwhm_plot}
\end{figure*}

}}

\newpage


\bibliography{citations}{}
\bibliographystyle{aasjournal}



\end{document}